\documentclass[preprint2]{emulateapj}

\newcommand{\Msun}{M\ensuremath{_{\odot}}}
\newcommand{\Msunyr}{M\ensuremath{_{\odot}}yr\ensuremath{^{-1}}}

\newcommand{\Lya}{Ly\ensuremath{\alpha}\,}
\newcommand{\kms}{{\rm km\,s}$^{-1}$}
\def\lsim{\mathrel{\rlap{\lower4pt\hbox{\hskip1pt$\sim$}}
    \raise1pt\hbox{$<$}}}                
\def\gsim{\mathrel{\rlap{\lower4pt\hbox{\hskip1pt$\sim$}}
    \raise1pt\hbox{$>$}}}                

\usepackage[small,normal,bf,up]{caption}
\renewcommand{\captionfont}{\small}

\usepackage{natbib,natbibspacing}

\shorttitle{The Mean UV Spectrum of $z\sim4$ LBGs}

\shortauthors{Jones, Stark, \& Ellis}

\begin{document}


\title[The Mean UV Spectrum of $z\sim4$ LBGs]{Keck Spectroscopy of Faint $3<z<7$ Lyman Break Galaxies: 
III. The Mean Ultraviolet Spectrum at $z\simeq4$}

\author{Tucker Jones\altaffilmark{1}, Daniel P. Stark\altaffilmark{2,3,4}, Richard S. Ellis\altaffilmark{1}
}

\altaffiltext{1}{Astronomy Department, California Institute of Technology, MC249-17, Pasadena, CA 91125, USA}
\altaffiltext{2}{Institute of Astronomy, Cambridge CB3 0HA, UK}
\altaffiltext{3}{Steward Observatory, University of Arizona, Tucson, AZ 85721}
\altaffiltext{4}{Hubble Fellow}



\begin{abstract}

We present and discuss the mean rest-frame ultraviolet spectrum for a sample of 81 Lyman Break Galaxies (LBGs) selected to be B-band
dropouts with a mean redshift of $z=3.9$ and apparent magnitudes $z_{AB}<26$. Most of the individual spectra are drawn from our ongoing survey 
in the GOODS fields with the Keck DEIMOS spectrograph described in earlier papers in the series, and we have augmented our sample with 
published data taken with FORS2 on the VLT. In general we find similar trends in the spectral diagnostics to those 
found in the earlier, more extensive survey of LBGs at $z=3$ undertaken by 
\cite{Shapley03}. Specifically, we find low-ionization absorption lines 
which trace the presence of neutral outflowing gas are weaker in galaxies with stronger \Lya emission, bluer UV spectral slopes, lower stellar 
masses, lower UV luminosities, and smaller half-light radii. 
This is consistent with a physical picture whereby star formation drives outflows of neutral gas which scatters \Lya and gives rise to strong low-ionization absorption lines, while increasing the stellar mass, size, metallicity, and dust content of galaxies.
Typical galaxies are thus expected to have stronger \Lya emission and weaker low-ionization absorption at earlier times (higher redshifts). 
Indeed, our mean spectrum at $z\simeq4$ shows somewhat weaker low-ionization
absorption lines than at $z=3$ and available data at high redshift
demonstrate that this evolutionary trend continues.
Although the total absorption by low-ionization transitions weakens at high redshift, the fine structure emission lines 
are stronger suggesting a greater concentration of neutral gas at small galactocentric radius ($\lsim 5$ kpc). 
In conjunction with earlier results
from our spectroscopic survey which demonstrated an increased fraction of LBGs with \Lya emission at higher redshift, we argue that the reduced 
low-ionization absorption is likely caused by a decrease in the covering fraction and/or velocity range of outflowing neutral gas at earlier epochs.
At present we cannot distinguish
between differences
in the covering fraction, outflow kinematics and geometry as the
underlying 
cause of this interesting evolution. However, our continuing survey
will enable us to extend these
diagnostics more reliably to higher redshift and determine the
implications
for the escape fraction of ionizing photons which governs the role of
early galaxies
in cosmic reionization.

\end{abstract}

\section{Introduction}

Considerable progress has been made over the past decade in charting the demographics of high redshift galaxies. 
Multi-wavelength surveys have defined the luminosity functions of UV and sub-mm selected star-forming sources \citep{Reddy09, Wardlow11}
as well as the coeval population of quiescent massive red galaxies 
\citep{Brammer11}. Spitzer data has revealed the time-dependent {\it stellar mass density} - a complementary quantity 
which represents the integral of the past star formation activity (e.g. \citealt{Stark09}).  Through these surveys a well-defined picture 
of the history of star formation and mass assembly over $0<z<6$ has been empirically determined \citep{Hopkins06,Ellis08,Robertson10}.
The redshift range $2<z<3$ corresponds to the peak of star formation activity where the 
Hubble sequence starts to emerge, and the earlier era corresponding to $3.5<z<5$ is an even more formative one where 
mass assembly was particularly rapid. 

Intermediate dispersion spectroscopy of carefully-selected Lyman break galaxies (LBGs) has been particularly important
in defining population trends that cannot be identified from photometric data alone. A very influential study at $z\simeq3$ was 
undertaken by \cite{Shapley03} who used composite Keck LRIS spectra of various subsets of nearly 1000 LBGs to 
examine the role of hot stars, H{\sc ii} regions and dust obscuration, as well as to measure the
outflow kinematics and absorption line properties of neutral and ionized gas. Composite spectra are particularly
useful for measuring weak lines which cannot be studied in detail for individual objects.
Through these careful studies, a detailed picture of the mass-dependent evolution
of LBGs has emerged (see \citealt{Shapley11} for a recent review).

In earlier papers in this series (\citealt{Stark10}, hereafter Paper I; \citealt{Stark11}, hereafter Paper II), we introduced an
equivalent spectroscopic survey of LBGs selected from a photometric catalog of more distant LBGs with $3<z<7$ in 
the Great Observatories Origins Deep Survey (GOODS) fields \citep{Giavalisco04, Stark09}. Whereas the \cite{Shapley03}
study targetted the study of LBGs close to the peak of activity in the overall cosmic star formation history, 
this earlier period corresponds to a less well-studied era when the rate of mass assembly is particularly rapid. From photometric
data alone, \cite{Stark09} deduced some significant changes in the characteristics of star formation at $z \simeq 4-6$
compared to later times, for example a shorter timescale of activity ($\simeq300$ Myr). We considered it crucial to understand 
these changes in LBG properties if these galaxies are to be used as probes of cosmic reionization at higher redshifts.

At the time of writing, our Keck survey is continuing with increasing emphasis at high redshift. Paper I presented the first 
substantial results from a survey of LBGs at $3.5<z<6$ observed with the Keck/DEIMOS spectrograph.
Paper II augmented this data with a further sample following more ambitious exposures focusing primarily on 
$z\simeq6$ LBGs. Incorporating a sample of ESO VLT spectra, retrospectively selected using similar photometric criteria 
as those for the Keck sample from the FORS2 study of \cite{Vanzella05, Vanzella06, Vanzella08, Vanzella09}, the current dataset amounts to a sample of 546 galaxies over the redshift range $3.5<z<6.3$. 

Our earlier papers in this series concentrated primarily on the rate of occurrence of Lyman $\alpha$ (\Lya) emission in our 
spectra (the ``\Lya fraction''). The overall goal was to understand the significantly different evolutionary trends in the luminosity 
functions of LBGs and narrow-band selected \Lya emitters (LAEs, \citealt{Ouchi08}) prior to the use of the Ly$\alpha$ fraction as
a test of when reionization ended \citep{Schenker11}. Paper I confirmed a result found by \cite{Shapley03} at lower redshift, 
namely that \Lya emission is more frequent in lower luminosity LBGs rising to a high proportion $\simeq 50$\%
at $M_{UV}=-19$. More importantly, the Ly$\alpha$ fraction was found to rise modestly with redshift over $3<z<6$.
By correlating the visibility of Ly$\alpha$ emission with UV continuum slopes derived from the HST photometry, it was argued 
that these trends in the visibility of line emission most probably arise from different amounts of dust obscuration. 
Reduced dust extinction in lower luminosity LBGs and those at higher redshift has also been deduced from studies of 
larger photometric samples  (\citealt{Bouwens09}, see also \citealt{Reddy09}). 

Paper I also discussed the possibility that the covering fraction of hydrogen may be lower in low-luminosity LBGs.
Strong \Lya emission in luminous LBGs is often associated with low equivalent width interstellar absorption lines
arising from a non-uniform covering fraction of neutral hydrogen \citep{Quider09, Shapley03}.
This trend suggests that the high \Lya fraction in faint LBGs is partially due to a lower covering fraction,
which would imply that Lyman continuum photons may more easily escape from intrinsically faint $z\simeq4-6$ galaxies.
Such a result would have great importance in understanding the role of $z>7$ galaxies in maintaining cosmic
reionization \citep{Robertson10}.

The present paper represents our first analysis of the spectral properties of $z\simeq4-5$ LBGs derived from composite
spectra in the manner pioneered at $z\simeq3$ by \cite{Shapley03}. The large database now amassed
following the campaigns at Keck and the VLT makes a similar study now practical in the redshift range where there is 
evidence of increased short-term star formation and the mass assembly rate is particularly rapid. Via detailed studies of 
low-ionization absorption line and emission line profiles, we aim to examine possible changes in the kinematics and
covering fraction of neutral gas, which affects the strength of \Lya and the escape fraction of ionizing photons,
as we approach the reionization era. As our redshift survey continues, in this 
paper we focus on a sample of galaxies with $3.5<z<4.5$, selected as B-dropout LBGs. Combining our Keck sample with 
data from the VLT (see Paper I for details), herein we examine the spectral features and trends in composite spectra drawn from a sample of 131 galaxies.

A plan of the paper follows. We briefly review the spectroscopic observations and their data reduction in \S2; much of the
relevant discussion is contained in Paper I. In \S3 we describe the selection of individual spectra that we consider appropriate for
forming the composite mean spectrum at $z\simeq4$ and the associated sources of uncertainty. \S4 examines the mean
spectrum in detail and introduces the various diagnostic features in the context of a physical model for LBGs of different
masses and star-formation rates. In \S5 we compare spectroscopic trends grouped by observable properties such as mass and luminosity with those
found at $z\simeq3$ by \cite{Shapley03}. In \S6 we discuss those trends which appear to be 
redshift-dependent, discussing implications for the role of early star-forming galaxies in cosmic reionization.
Finally we summarize our results in \S7.

Throughout this paper, we adopt a flat $\Lambda$CDM cosmology with $\Omega_{\Lambda} = 0.7$, $\Omega_M = 0.3$,
and $H_0= 70$ $h_{70}$ \kms Mpc$^{-1}$. All magnitudes in this paper are quoted in the AB system \citep{Oke74}.

\section{Observations and Data Reduction}

The rationale and procedures used to undertake our spectroscopic survey of LBGs over $3<z<7$ were introduced in
detail in Paper I and the interested reader is referred to that paper for further detail. Here we recount only the basic
details. Our target LBGs were selected as $B$, $V$ or $i'$--band ``dropouts'' based on deep
photometry in the two GOODS fields.
The photometric catalog used in our analysis will be described in detail in
Stark et al. (2011, in preparation).  The selection and photometric approach
is largely similar to that described in \cite{Stark09} and Paper I,
but we highlight two key updates.   First the selection is performed
on the v2 GOODS ACS catalogs.   Second, we utilize deep ground-based near-IR
imaging in GOODS-N obtained from WIRCAM on CFHT \citep{Wang10} and deep
HST Wide Field Camera 3 / IR imaging of GOODS-S from CANDELS \citep{Grogin11, Koekemoer11}.

\subsection{Keck/DEIMOS}

The majority of spectra we present are taken from an ongoing survey with the DEep Imaging Multi-Object Spectrograph 
(DEIMOS; \citealt{Faber03}) on the Keck II telescope.  In this paper we have used observations taken in 2008 April and 
2009 March (masks GN081, GN082, GN083, GN094, and GN095 in Paper I), which targeted a total of 261 $B$-drop 
and 88 $V$-drop galaxies. These data were taken with the 600 lines\,mm$^{-1}$ grating, covering the wavelength range 
4850--10150 \AA\, with a resolution of $\simeq 3.5$ \AA. We do not use data at wavelengths $\lambda > 9200$ \AA\, which are
affected by strong and variable absorption by atmospheric water vapor. The seeing was typically 0\farcs8 and ranged 
between 0\farcs5 and 1\farcs0 during the observations. 

All data were reduced and calibrated using a modified version of the IDL pipeline {\sc spec2d}, developed specifically for 
DEIMOS by the DEEP2 survey team \citep{Davis03}. The data were reduced as described in Paper I, with the addition of two important 
modifications. First, the continuum traces of all target galaxies and other objects occupying the same slit were carefully 
masked to exclude object flux from the sky background model. 
Second, the b-spline fit to the sky background was modified to include a 2nd-order polynomial fit to the spatial 
dimension. These modifications significantly improved the sky subtraction, particularly at long wavengths $\lambda > 7000$ \AA\ 
where bright sky lines can be problematic.

\begin{figure*}
\includegraphics[width=\textwidth]{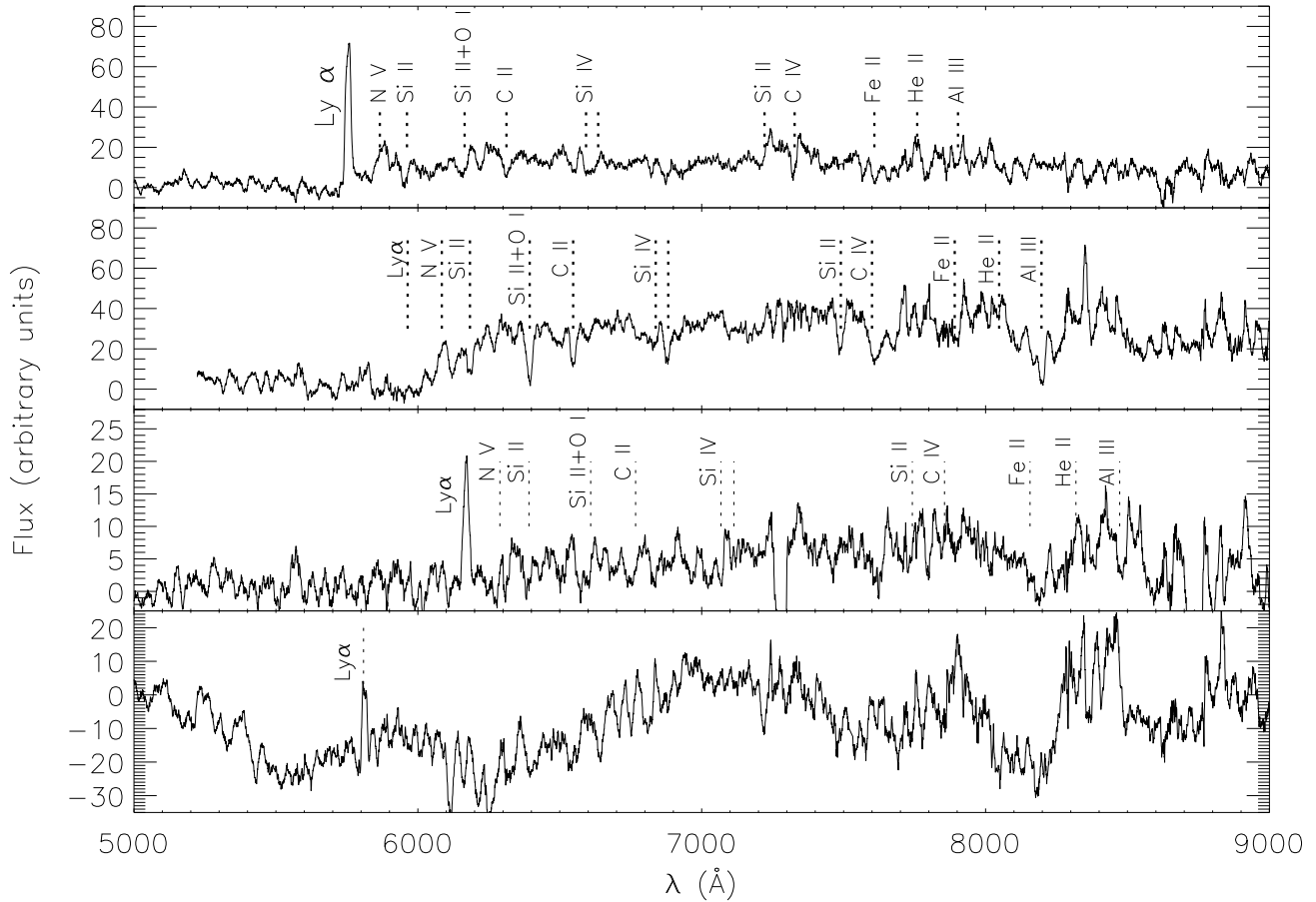}
\caption{\label{fig:spectra} Examples of individual DEIMOS spectra for $3.5<z<4.5$ LBGs. From top to 
bottom: a galaxy at $z=3.73$ with \Lya emission and weak interstellar absorption features; a galaxy at $z=3.906$ with \Lya and 
interstellar absorption; a faint galaxy at $z=4.07$ with \Lya emission and no detectable absorption lines; and  galaxy at $z=3.778$ with \Lya emission whose spectrum is contaminated by scattered light and detector defects. 
The top three spectra are included in stacking analyses while the bottom spectrum is excluded based on poor data quality.
The continuum signal-to-noise in each spectrum degrades noticeably at wavelengths $\lambda > 7000$ \AA\, due to increased OH sky emission.}
\end{figure*}

The reduced one- and two-dimensional spectra were visually inspected using the IDL program {\sc SpecPro} \citep{Masters11}. 
Spectra which suffered from poor data quality were excluded from further analysis. The exclusions included unacceptable amounts
of scattered light within the detector mosaic, defective CCD columns, contamination from bright nearby sources, or poor sky 
subtraction, for example arising from the location of the slit on the detector. A few low-redshift interlopers were also identified 
and excluded from further analysis. For the remaining spectra, redshifts were measured from either \Lya emission (where present) 
or interstellar absorption lines. Galaxies identified as hosting strong AGN based on the presence of C{\sc iv} or other strong emission lines 
were excluded. We identified 5 stars, 2 dusty low-redshift galaxies ($z\sim0.5$), 4 AGN, and 134 star-forming galaxies with secure redshifts $z>3$. After rejecting poor data, the final sample consists of 94 high-quality spectra with accurate redshifts. Examples of high redshift dropout 
spectra are shown in Figure~\ref{fig:spectra}.

\subsection{Archival VLT/FORS2 Spectroscopy in GOODS-S}

To augment our sample of high redshift spectra, we have also made use of data from the FORS2 program of \cite{Vanzella05, Vanzella06, Vanzella08, Vanzella09} 
which targeted dropouts in the GOODS-S field. The characteristics of that survey in terms of resolution and spectral coverage are very similar to 
that undertaken at Keck with DEIMOS and details can be found in Paper I. Using the coordinates provided in the published FORS2 database
we queried the version 2.0 ACS catalogs for GOODS-S and undertook our own photometric measures and dropout selection criteria in an 
identical fashion to that used for our Keck survey. The magnitude distribution of the FORS2 sample is generally weighted towards sources brighter
than those in the overall Keck survey, but for the purposes of constructing the mean spectra discussed in this paper, the bulk of the individual
spectra are of comparable brightness.

\subsection{Redshift measurements}\label{sec:redshifts}

Care is needed in deriving accurate systemic redshifts from rest-frame UV spectra since the strongest features trace the kinematics of outflowing 
gas rather than that of the stars. Stellar absorption lines are usually too faint to be measured precisely given the signal to noise of the spectra.  
Typically the only features detected in individual spectra are \Lya and strong interstellar absorption lines such as Si{\sc ii}$\lambda$1260, O{\sc i}$\lambda$1302+Si{\sc ii}$\lambda$1304, C{\sc ii}$\lambda$1334, Si{\sc iv}$\lambda\lambda$1393,1402, Si{\sc ii}$\lambda$1526, and C{\sc iv}$\lambda\lambda$
1548,1550 (Figure~\ref{fig:spectra}). Absorption by outflowing gas results in a blueshift of interstellar absorption lines. Outflowing neutral hydrogen along the
line of sight leads to a \Lya profile which displays very broad ($\gsim 1000$ \kms) blueshifted absorption and net redshifted emission. The high-ionization 
Si{\sc iv} and C{\sc iv} lines also arise in P-Cygni stellar winds with broad blueshifted absorption. The magnitude of these offsets has been well-quantified for
star-forming galaxies at $z\simeq2.3$, and is typically $-200$ \kms for interstellar absorption lines and $+500$ \kms for \Lya emission \citep{Steidel10}.

To determine accurate redshifts for making composite spectra, we restrict our sample to those with redshifts measured from either \Lya emission 
($z_{\Lya}$) or low-ionization interstellar absorption lines ($z_{IS}$). Although, as discussed above, these are not at the systemic redshift, we follow
well-established techniques to correct for the typical offsets \citep{Steidel10}. We do not use redshifts based on \Lya absorption or high-ionization lines 
(Si{\sc iv} and C{\sc iv}) because of the complex and variable blueshifts of these features with respect to the systemic redshift. We define $z_{\Lya}$ as the
centroid of the emission line and consider only spectra in which the line is detected at $>5\sigma$ significance. Interstellar absorption line redshifts require
careful treatment to avoid spurious identification of sky line residuals. We consider only the low-ionization Si{\sc ii}$\lambda$1260, O{\sc i}$\lambda
$1302+Si{\sc ii}$\lambda$1304, and C{\sc ii}$\lambda$1334 features which are typically found in the highest signal-to-noise regions of our spectra. Absorption
features at longer wavelengths are less reliable due to the higher density of strong night sky lines, while shorter wavelength transitions are lost in the \Lya forest.
To measure absorption line redshifts, we first estimate the redshift from \Lya (either in emission or absorption). We then fit Gaussian profiles to the spectrum near the
expected position of the three features above. We require the best fit of all three lines to be consistent to within $\pm 500$ \kms with a combined significance of
$>5\sigma$. If these conditions are met, we define $z_{IS}$ as the weighted mean redshift of the three interstellar features.

\begin{figure}
\includegraphics[width=0.5\textwidth]{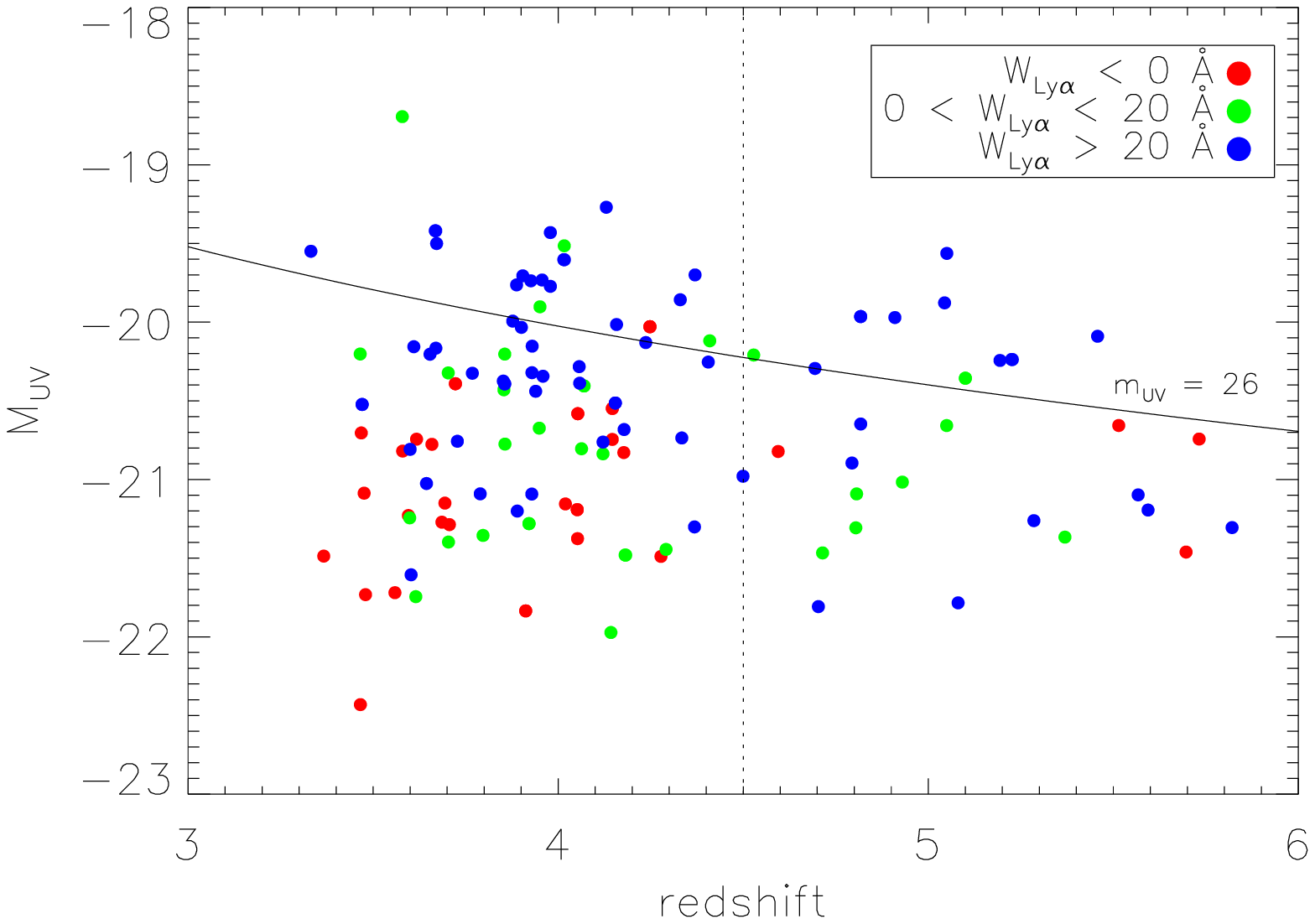}
\caption{\label{fig:sample} Redshifts and UV luminosities for the sample of 131 galaxies with suitably accurate redshifts measured either from \Lya or strong
interstellar absorption lines. Points are color-coded according to the equivalent width of \Lya, W$_{\Lya}$. Galaxies brighter than an apparent magnitude
$z'_{AB} < 26.0$ (solid line) typically have well-detected continua suitable for forming a composite spectrum.
Galaxies with redshift $z<4.5$ (dashed line) are used for the composite spectra discussed in \S\ref{sec:composite}-\S\ref{sec:trends}, while \S\ref{sec:cgm} includes galaxies at all redshifts shown here.
}
\end{figure}

With the criteria above, suitable redshifts are available for a total of 131 high-quality DEIMOS and FORS2 spectra. 91 redshifts are based on measures of
$z_{\Lya}$ emission only, 31 have $z_{IS}$ only, and 9 have both $z_{\Lya}$ and $z_{IS}$. The distribution of redshifts and absolute UV magnitudes for this
sample is shown in Figure~\ref{fig:sample}. Unless stated otherwise, further analysis in this paper is restricted to the 81 sources with redshift $z<4.5$ and
apparent magnitude $z' _{AB}< 26.0$. We applied this additional magnitude criterion in order to ensure a well-defined continuum signal/noise in each individual spectrum.
53 of these 81 sources are drawn from the Keck survey and 28 from FORS2.

\subsection{Sample bias}\label{sec:sample_bias}

While the sample is 90\% complete for dropout selected galaxies
to an apparent magnitude $z_{AB}=25$, at fainter magnitudes
\Lya emission or strong interstellar absorption features
are required for reliable redshifts. This results in a bias towards
stronger low-ionization absorption lines for the fraction
($\simeq35$\%) without \Lya. We can quantify this bias through the
detectability of the average low-ionization absorption line strength shown in the
composite spectrum in Figure~\ref{fig:composite} (see \S\ref{sec:composite}). 
We find that we can measure
redshifts for this average absorption line strength with 90\%
completeness at $z_{AB}=24.7$, declining through 50\% at $z_{AB}=25.2$ to zero
at $z_{AB}>25.5$. Of the galaxies with detected absorption line
redshifts, those with $z_{AB}>25.2$ have line strengths only
10\% stronger than for those with $z_{AB}<25.2$. Clearly this
is a small effect.

The bias toward stronger \Lya emission for fainter galaxies
is manifest in Figure~\ref{fig:sample} where a paucity of objects with
$W_{\Lya}<20$
\AA\ can be seen at faint magnitudes. This bias was fully quantified
in Paper I using Monte Carlo simulations. For galaxies at $3.5<z<4.5$, a sample
completeness of 95\% is reached at $W_{\Lya} \gsim 20$ \AA\, for $z'_{AB} = 26$, and $W_{\Lya} \gsim 7$ \AA\, for $z'_{AB} = 25$. 
For the sample presented in this paper, the least biased subset is that with strong \Lya emission, followed by that with bright apparent magnitudes $z'_{AB}<25$.

A final issue in considering composite spectra is that these are comprised of individual spectra across $3.5<z<4.5$ with different rest-frame
wavelength ranges. As our spectra generally cover the wavelength range $5000 < \lambda < 9200$ \AA, galaxies at $z=3.5$ contribute to the rest frame 
$1100-2050$ \AA, while those at $z=4.5$ contribute to $900-1650$ \AA. At longer rest-frame wavelengths the composite spectrum is therefore largely
contributed by galaxies at lower redshift. In addition there are wavelength-dependent sources of noise discussed in \S\ref{error_spectrum}. 
In summary, the data used to construct the composite spectrum in Figure~\ref{fig:composite} correspond to a mean redshift $z = 3.98$ at $\lambda_{rest} = 1100$ \AA, $z = 3.95$ at 1200 \AA, $z = 3.90$ at 1500 \AA, and $z=3.82$ at 1650 \AA. For the wavelength range of interest in this work, 
the redshift bias $\Delta z < 0.15$ is not particularly troublesome.

\section{Composite Spectra}\label{sec:composite}

We now turn to the presentation of the composite spectra. We need to account for the difference between the redshifts determined using \Lya on the one
hand and the low-ionization interstellar lines on the other hand and the systemic redshift prior to co-addition. We also seek to understand the signal/noise
of the composite in terms of the statistical uncertainties and the variance among the individual spectra used to construct the composite.

Composite spectra are constructed by shifting the individual spectra into the rest frame according to a deduced systemic redshift and then averaging the
set. In general terms we will first identify a subsample based on their observable properties. Each spectrum in the sample is shifted to the adopted rest frame 
and interpolated to a common wavelength scale with a dispersion of 0.12 \AA. All spectra are normalized to have a median $f_{\nu} = 1$ in the range
1250--1500  \AA. Spectra taken with DEIMOS are smoothed to a resolution of 1.9 \AA to match the lower resolution of FORS2 data. The spectra are then 
averaged at each wavelength using a $\sigma$-clipped mean to reject outliers arising from sky subtraction residuals and cosmetic defects. 
An equal number of positive and negative outliers are rejected at each wavelength, totaling at most 30\% of the data. The remaining data are averaged
with an arithmetic mean.

Uncertainty in a composite spectrum will arise from both the finite signal to noise and the variance of the individual galaxies. For example, the variance 
in \Lya equivalent widths in our sample is much greater than the uncertainty measured in the individual spectra. It is especially important to quantify the sample
variance for weak features that are generally not detected in individual spectra. We account for sample variance with a bootstrap technique. For each composite
spectrum we create 100 alternate composites using the same number of spectra but drawn at random from the parent sample. Each alternate has an average 63\% of the sample represented with 37\% duplicates. Every measurement made on the composite spectrum is repeated for each of the 100 alternates. 
We then take the measurement error to be the standard deviation of the 100 alternate measurements, which reflects both the sample variance and finite signal 
to noise.

As discussed in \S2.3 the most challenging issue is to determine the systemic redshift prior to shifting to the rest-frame. Here we follow the approach used
by \cite{Shapley03}. As a first approximation we use the value of $z_{\Lya}$ (where available) to construct a composite spectrum. This enables us to 
locate the stellar photospheric line C{\sc iii}$\lambda$1176 in the composite where we detect a velocity difference of $-330$ \kms\, with respect to \Lya. 
We can thus infer that \Lya emission in our sample is redshifted on average by $\Delta v_{\Lya} = 330$ \kms. In a similar fashion, stacking spectra 
using the redshift $z_{IS}$ based on low-ionization interstellar absorption results in a detection of C{\sc iii}$\lambda$1176 with a velocity offset 
of $+190$ \kms. For comparison, at $z\simeq$3  Steidel et al (2010) find $\left< \Delta v_{\Lya} \right> = +445$ \kms\, and $\left< \Delta v_{IS} \right> = -164$
 \kms. 
To construct composite spectra, we use either $z_{\Lya}$ shifted by $-330$ \kms or $z_{IS}$ shifted by $+190$ \kms to approximate the systemic redshift of each galaxy.
We use the \Lya-based redshift when available since it is typically determined with greater precision than $z_{IS}$. Figure~\ref{fig:composite} 
 shows the composite spectrum of 81 galaxies in our sample with $3.5<z<4.5$ and apparent magnitude $z'_{AB} < 26.0$ using this method.

\subsection{Uncertainties in the Systemic Redshift}

\begin{figure*}
\includegraphics[width=\textwidth]{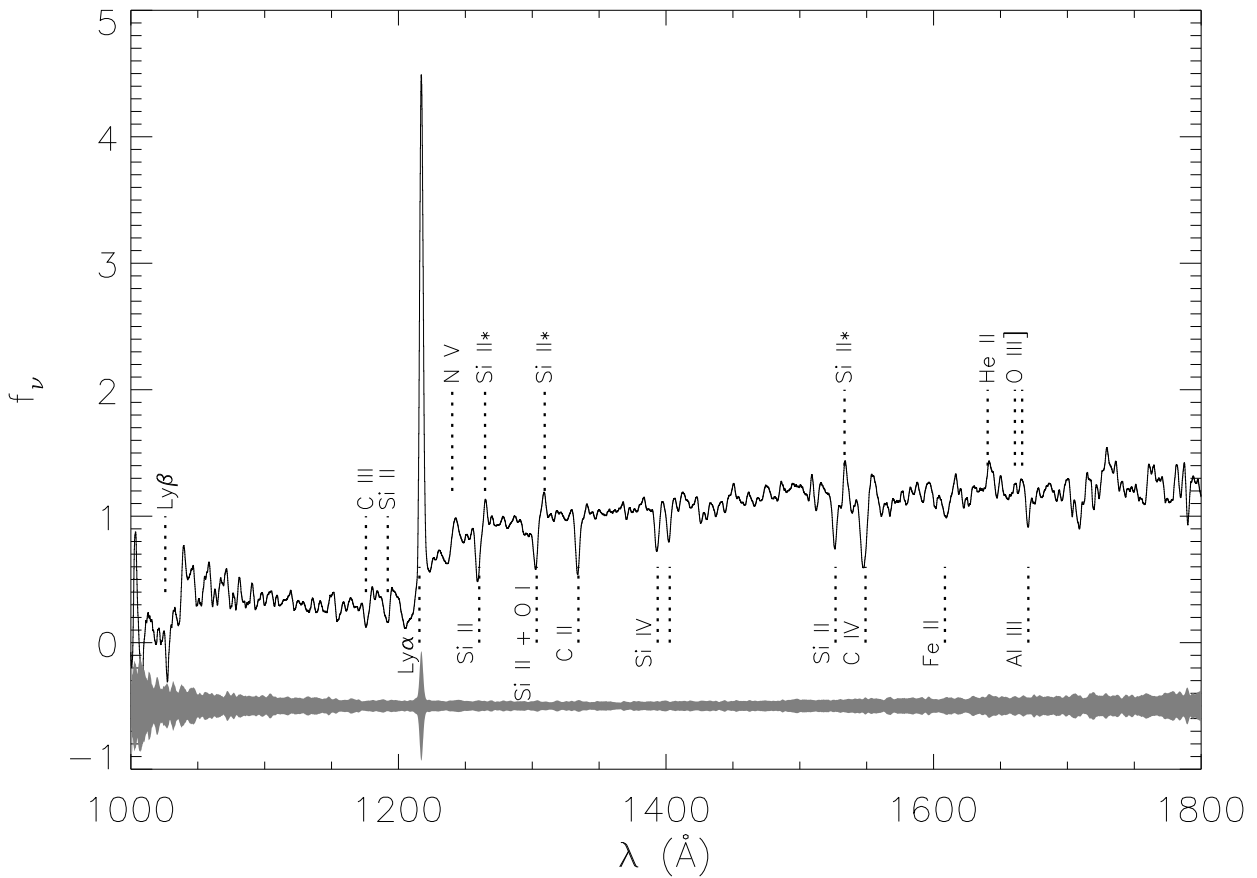}
\caption{\label{fig:composite} The composite spectrum of  81 galaxies in our sample with $3.5<z<4.5$ and apparent magnitude $z'_{AB} < 26$. The 
effective mean redshift for the sample averaged over wavelength is $\overline{z}=3.9$. The strongest spectral features are labelled. The gray filled region 
shows the $\pm 1 \sigma$ error at each point, determined from the scatter of individual spectra used to create the composite. The error spectrum peak 
at 1216 \AA\, is due to large scatter in the intrinsic distribution of \Lya equivalent widths. The error is lowest at $\sim 1300-1500$ \AA\, where the 
continuum signal to noise ratio is $\simeq 30$.  The error increases at shorter wavelengths where the instrument throughput is 
lower, and at longer wavelengths where sky emission is much stronger.}
\end{figure*}

A natural concern is the extent to which these applied shifts might vary within the sample used to make the composite. This can be estimated from
 observations of higher signal to noise from spectra taken at lower redshift. \cite{Steidel10} quantify the offset between $z_{IS}$, $z_{\Lya}$, and the 
 systemic redshift $z_{H\alpha}$ in a sample of 89 galaxies at $z\simeq2.3$. They find velocity offsets $\Delta v_{IS} = -170 \pm 115$ \kms\, and 
 $\Delta v_{\Lya} = 485 \pm 185$ \kms relative to H$\alpha$. Assuming this is representative of our data, the uncertainty in our systemic redshift 
 is therefore likely to be $\sigma(v)\sim150$ \kms. An upper limit on $\sigma(v)$ can be estimated from the width of spectral lines in the composite
 spectrum. In particular the rest-wavelength of the stellar line C{\sc iii}$\lambda$1176 in the composite provides a valuable measurement of the 
 average offset from the systemic velocity and its width provides an upper limit on the effective spectral resolution. We measure a systemic velocity of 
 $21\pm101$ \kms\, in the composite spectrum (Figure~\ref{fig:composite}) and a FWHM $= 520$ \kms\, (deconvolved from the instrumental resolution $\simeq450$ \kms). The uncertainty in the adopted redshifts about the true systemic stellar value is therefore $\leq 520$ \kms\, FWHM or 
 equivalently $\sigma(z) \leq 220$ \kms, comparable in fact to what was achieved for individual spectra at lower redshift by \cite{Steidel10}.

\begin{figure*}
\includegraphics[width=\textwidth]{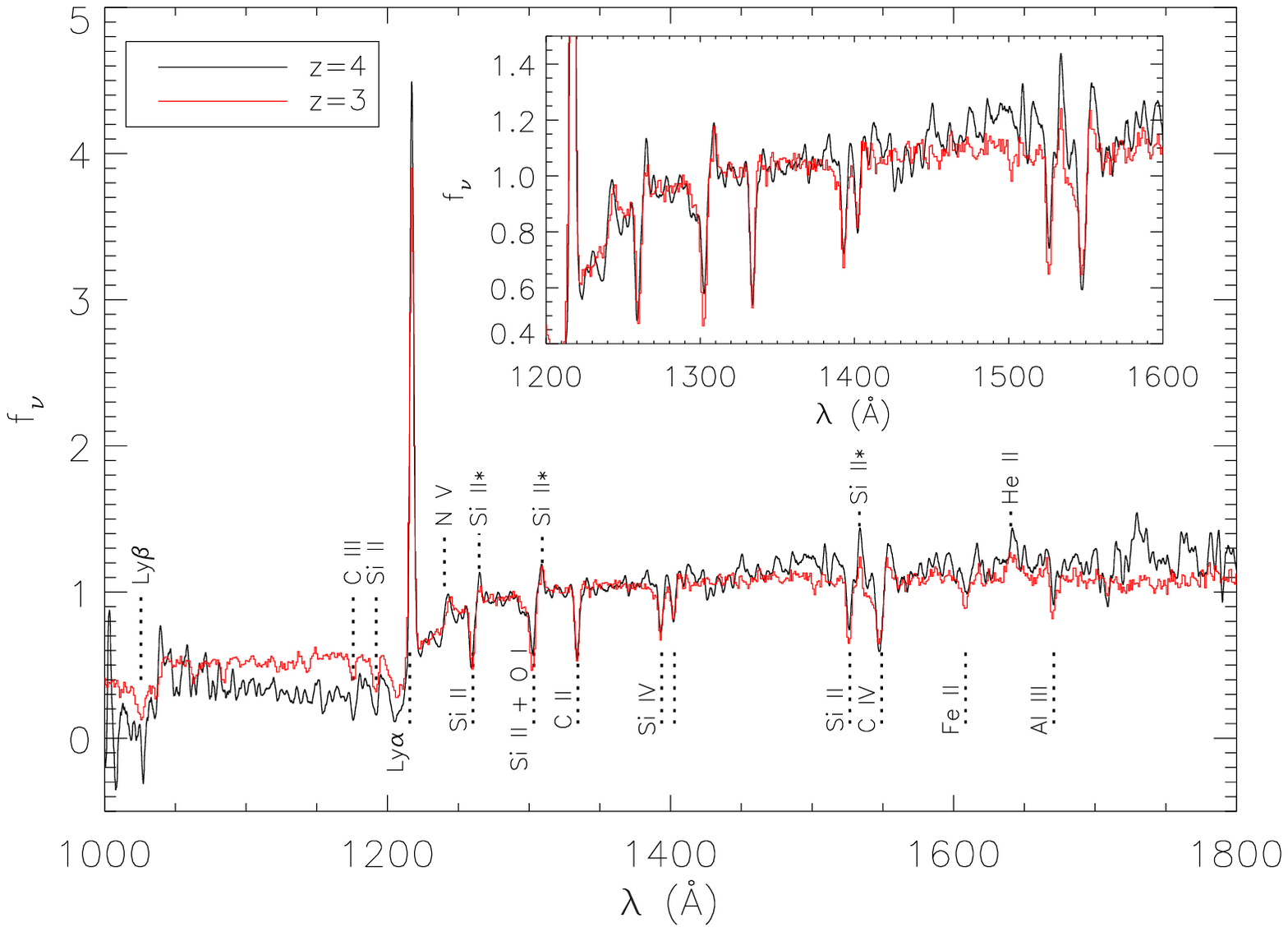}
\caption{\label{fig:composite2} Composite spectrum of all 81 galaxies in our sample with $z<4.5$ and apparent magnitude $z' < 26$, compared to the composite spectrum of 811 LBGs at $z=3$ presented in \cite{Shapley03}. The higher redshift sample has a much stronger \Lya forest break, slightly redder UV spectral slope, slightly stronger \Lya emission, and weaker absorption lines. The inset shows a zoom-in of the region from $1200-1600$ \AA, which contains most of the absorption lines of interest in this paper.}
\end{figure*}

\subsection{Error spectrum}\label{error_spectrum}

Figure~\ref{fig:composite} shows the composite spectrum of our sample as well as the $1\sigma$ error spectrum derived using the bootstrap technique
discussed above. The error at each pixel is calculated as the standard deviation of all averaged data points (excluding outliers), divided by the square 
root of the number of data points. There are several wavelength-dependent factors contributing to the error spectrum in addition to the finite signal-to-noise 
of individual spectra. One factor is the intrinsic sample variance, seen clearly as a noise spike at the position of \Lya, and also evident for 
C{\sc ii}$\lambda$1334 and other absorption features. Another is the decreased instrument throughput at $\lambda \lsim 6000$ \AA, leading to 
higher noise at shorter wavelengths. Similarly, stronger sky line emission causes increasing noise at longer wavelengths. Finally, the number of contributing
spectra peaks at rest-frame  $\lambda \simeq 1150-1350$ \AA\,, with increased noise at higher and lower wavelengths where fewer spectra are included. 
Our observed wavelength range $5000<\lambda < 9200$ \AA\, corresponds to a rest-frame $1000<\lambda<1800$ \AA\, at the mean redshift 
$z = 3.9$ of our magnitude-limited sample. Ultimately we acheive a signal to noise ratio in the continuum of $\geq10$ between the \Lya line and rest-frame 1800 \AA\, and $\simeq 5$ in the \Lya forest, with a peak S/N $\simeq30$ at 1350 \AA.

\section{Features in the Composite Spectrum}

We now discuss the composite spectrum (Figure~\ref{fig:composite}) in more detail, focusing on the strong spectral features at 1215--1550 \AA\, where we 
have the best signal to noise. In this wavelength range we detect \Lya, N{\sc v}$\lambda\lambda$1239,1243, Si{\sc ii}$\lambda$1260, Si{\sc ii}*$\lambda$1265,
O{\sc i}$\lambda$1302+Si{\sc ii}$\lambda$1304 (blended), Si{\sc ii}*$\lambda$1309, C{\sc ii}$\lambda$1334, Si{\sc iv}$\lambda\lambda$1394,1403, Si{\sc ii}
$\lambda$1527, Si{\sc ii}*$\lambda$1533, and C{\sc iv}$\lambda\lambda$1548,1550 at high significance. Our $z=3.9$ composite spectrum is very similar to 
the composite of $z=3$ LBGs presented in \cite{Shapley03}, which we show in Figure~\ref{fig:composite2} for comparison. 
The absolute magnitude distribution of our sample is broadly similar to that at $z=3$; both cover the range $-22<M_{UV}<-20$.
Our discussion below
follows closely that presented originally by \cite{Shapley03} but we are also interested in whether there are differences seen over the redshift range
$3<z<4.5$. We will discuss these possible evolutionary trends in \S5.

\subsection{Lyman Break Galaxies: a physical picture}\label{sec:physical_picture}

It is helpful to begin by describing a possible physical picture of LBGs based on many analyses of the extensive observations at 
$z\simeq3$ (see \citealt{Shapley11} for a recent review). Typical $\mathcal{L}$* LBGs at $z=3$ have ultraviolet half-light radii $\simeq 2.0$ kpc, stellar 
masses $\sim 3 \times 10^{10}$ \Msun, and star formation rates $\sim 50$ \Msunyr \citep{Bouwens04, Ferguson04, Shapley01}.
The star formation surface density is sufficient to drive ``superwinds'' of outflowing gas, similar to those seen in local galaxies where 
$\Sigma_{SFR} \gsim 0.1$ \Msunyr kpc$^{-2}$ \citep{Heckman02}. Indeed, blueshifted interstellar absorption lines confirm there are outflows with
typical velocities of $\simeq 150$ \kms , and in some cases as high as 800 \kms \citep{Shapley03, Pettini02, Quider09, Quider10}. However,
the physical origin of these outflows remains unclear (e.g. \citealt{Murray10}). It is thought these outflows produce an extended circumgalactic medium 
(CGM) of ejected material. Outflowing gas is found in both low and high ionization states (e.g. Si{\sc ii} and Si{\sc iv}). Low-ionization transitions such 
as Si{\sc ii} and C{\sc ii} are mostly associated with neutral hydrogen whereas high ionization lines occur in the fully ionized component. Based on 
trends in the strength of the various species, \cite{Shapley03} suggest a geometry in which discrete clouds of neutral gas are embedded in a 
halo of ionized gas. \cite{Steidel10} show that the neutral and ionized CGM components both extend to radii of at least 125 kpc.

\subsection{\Lya}\label{sec:lya}

\begin{figure}
\includegraphics[width=0.5\textwidth]{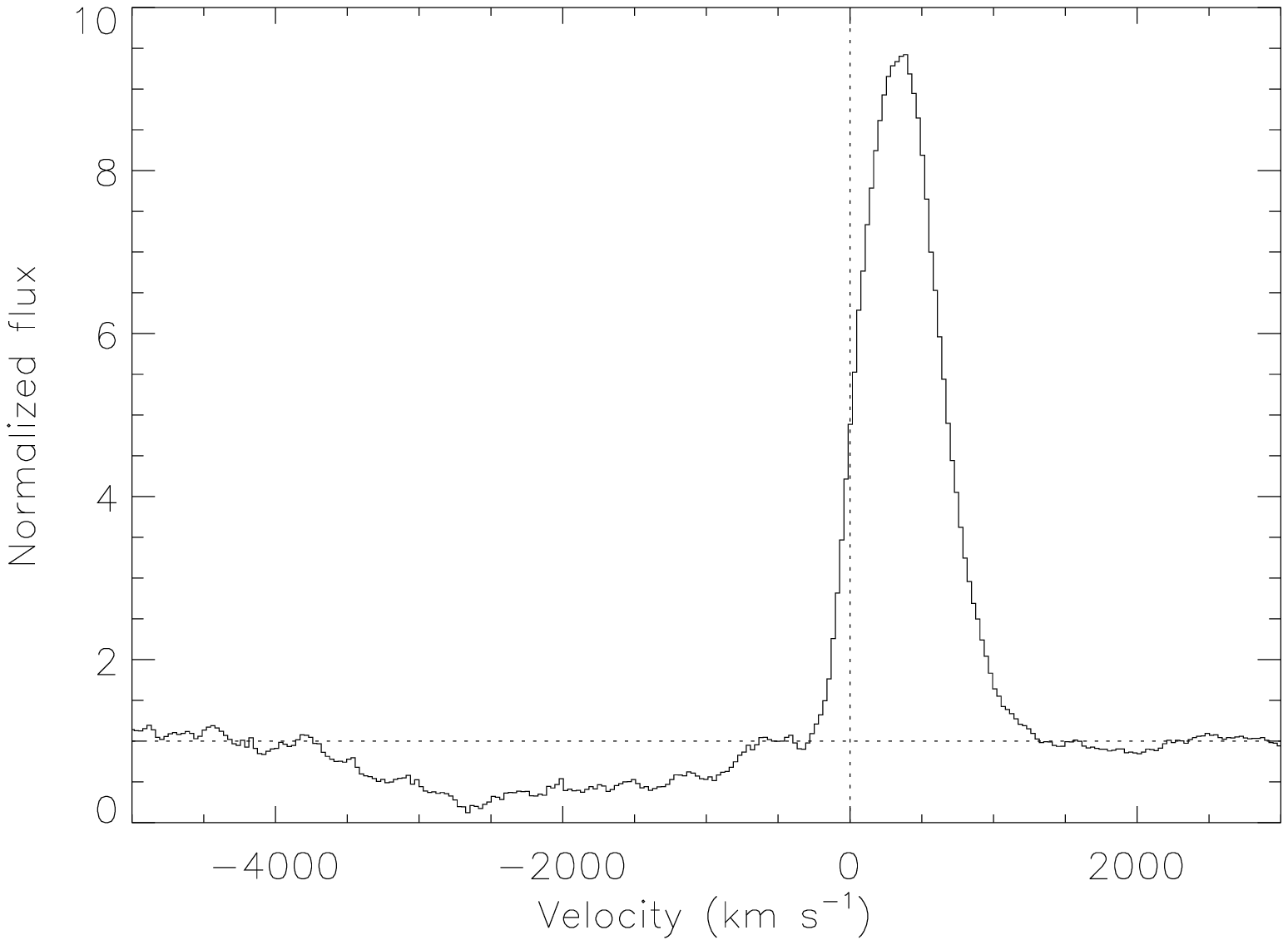}
\includegraphics[width=0.5\textwidth]{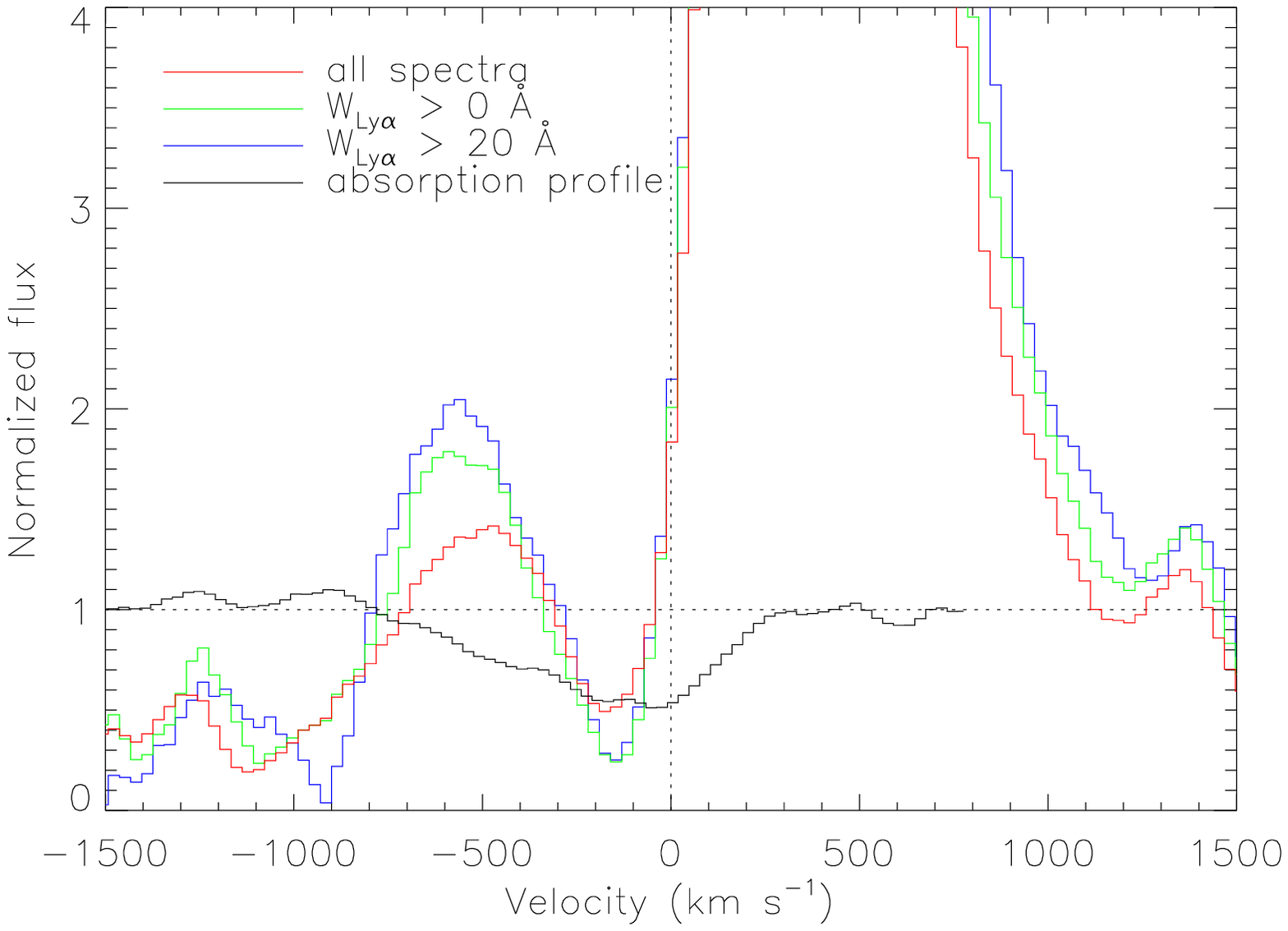}
\caption{\label{fig:lya_velocity} {\bf Top:} composite spectrum normalized to continuum flux levels, showing the velocity structure of \Lya. The line profile consists of a broad blueshifted absorption trough at $v \gsim -4000$ \kms and strong redshifted emission extending to $v=+1000$ \kms.
{\bf Bottom:} velocity profile of \Lya showing a blueshifted secondary peak. The composite spectra in this plot are constructed only from DEIMOS data, with spectral resolution $R\simeq2000$. The composites show a blueshifted \Lya emission peak centered at $v=-600$ \kms, corresponding to the maximum outflow velocity of neutral gas as seen in the average velocity profile of low-ionization absorption lines (black). Blueshifted \Lya emission arises from photons scattered at the leading edge of outflowing neutral CGM.}
\end{figure}

\Lya is the most prominent and diverse feature in our individual spectra. The line originates from hydrogen recombination in H II regions photoionized by 
massive stars and, as these stars dominate the adjacent continuum,  its {\it  intrinsic equivalent width} should be within the range $W_{\Lya} = 100-200$ \AA\, for nearly all stellar populations (e.g. \citealt{Forero11}).
Our composite spectrum reveals a complex line profile with strong absorption extending blueward to $v = -4000$ \kms, 
and redshifted emission to $v = +1000$ \kms (Figure~\ref{fig:lya_velocity}) with a peak offset of $v = +330$ \kms relative to the systemic velocity. 
The net equivalent width $W_{\Lya} = 21 \pm 3$ \AA\, is only $\lsim 20$\% of the expected intrinsic value. 

The form of this \Lya profile has been readily understood in terms of the physical picture discussed in \S\ref{sec:physical_picture}. \Lya emission produced 
at the systemic velocity can escape only along a line of sight free of neutral hydrogen or if it is shifted in velocity far from resonance. As photons escape
they encounter blueshifted clouds of partially neutral gas, which absorb and re-emit isotropically. Photons backscattered from neutral 
clouds at small radii will appear {\em redshifted} and have a higher chance of escaping. Additionally, photons scattered from neutral gas at the edge of 
the CGM can escape into the ionized intergalactic medium and will be observed as blueshifted emission \citep{Steidel10}. Indeed, 
Figure~\ref{fig:lya_velocity} shows such a weak blueshifted emission peak in the composite spectrum. On the other hand, the reduced equivalent width
compared to that expected intrinsically could be due to many effects including dust extinction, scattering at large radii where emission falls 
outside the spectroscopic slits and a non-zero escape fraction $f_{esc}$ of ionizing radiation.  Since these effects  cannot easily be disentangled, the 
absolute strength of the \Lya line must be interpreted with caution (Paper II, \citealt{Schenker11}).  

We can estimate the spatial extent of the neutral CGM (i.e. the radius at which escaping \Lya is last scattered) based on the strength of the diffuse blueshifted 
\Lya emission. The spatial profile of extended \Lya emission has been well-quantified at $z \simeq 2.6$ by \cite{Steidel11} who find that diffuse haloes 
are a generic property of star-forming galaxies leading to a total \Lya flux $\simeq 5 \times$ greater than that measured within the spectroscopic slits. 

Average \Lya surface brightness profiles are well-fit at large radii by an exponential form, viz.
\begin{equation}
\label{eq:lya}
\Sigma_{\Lya}(b) = \Sigma_0 \exp(b/b_l)
\end{equation}
where $b_l \simeq 25$ kpc. The blueshifted emission shown in Figure~\ref{fig:lya_velocity} has an equivalent width $W = 1.5 \pm 0.3$ \AA\, measured 
within a slit aperture of $\simeq1\farcs0 \times 1\farcs0$, corresponding to $7.1\times7.1$ kpc$^2$ at $z=4$. Assuming this approximates the peak 
value, we estimate $\Sigma_0 = 1.5\pm0.3 \text{ \AA}$ per 50 kpc$^2$ and integrating Equation~\ref{eq:lya} yields a total equivalent width of $W_{halo} = 2 \pi b_l^2 \Sigma_0$. Using the measured value of $\Sigma_0$, the scale length is given by
\begin{equation}
\label{eq:lya2}
b_l = (2.3 \pm 0.2) \sqrt{W_{halo}/\text{\AA}}\, \text{kpc}.
\end{equation}

Although the total \Lya flux is not measured, we can estimate its value from theoretical expectations as well as observations at lower redshift. The equivalent
width in Figure~\ref{fig:lya_velocity} is $W = 21.0$ \AA\, (excluding the blueshifted emission component), thus we can write $W_{halo} = W_{tot} - 21$ \AA. 
As discussed above, the {\em observed} $W_{tot}$ is diminished by dust and the escape of ionizing radiation. Indeed, \cite{Steidel11} measure $W_{tot} = 17-93$ \AA\, in various subsamples of their data, all lower than the expected intrinsic value, $W_{tot}$=100-200 \AA\ . Considering the composite DEIMOS spectra shown in Figure~\ref{fig:lya_velocity} and assuming an intrinsic $W_{tot} = 135$ \AA\ \citep{Forero11} we derive upper limits of $b_l < 20 \pm 2$ kpc for galaxies with $W_{\Lya} > 0$, and $b_l < 17 \pm 2$ kpc for those with $W_{\Lya} > 20$ \AA. This latter constraint is the most stringent, and also likely closest to the true value of $b_l$. \cite{Steidel11} find that their sample of \Lya emitters (LAEs, defined as $W_{\Lya} > 20$ \AA) has the highest $W_{tot}$ ($= 93$ \AA) and largest scale length ($b_l = 28.4$ kpc) of any sub-sample that they analyze. Our constraint of $b_l < 17$ kpc for the LAEs therefore suggests that the characteristic size of \Lya haloes may be smaller at $z=4$ than at $z=2-3$. Due to the large inherent uncertainties, direct measurements of low surface brightness \Lya emission are needed to confirm this possibility.

\subsection{Low-ionization metal transitions}\label{sec:lis}

According to the physical picture in \S\ref{sec:physical_picture}, absorption in low-ionization transitions occurs in both the interstellar medium and 
outflowing clouds of cool gas. The former is at the systemic velocity while the latter is blueshifted. Such transitions are generally saturated 
\citep{Shapley03, Pettini02} so the line depth at a given velocity provides a measure of the areal covering fraction $f_c$ of O and 
B stars by neutral gas along the line of sight. Specifically, the line profile is given by $f(v) = f_0 \cdot (1-f_c(v))$ where $f_0$ is the continuum flux. 
While typical LBGs are too faint for detailed line profiles, high-resolution spectra of a few bright lensed sources at $z=2-3$ \citep{Pettini02, Quider09, Quider10} have revealed absorption velocities ranging from $\sim-1000$ to $+500$ \kms\, with the highest covering fraction at $v\sim-200$ \kms. 
The mean low-ionization absorption line velocity offset in our composite is $v_{LIS} = -190$ \kms in good agreement with that at $z=2-3$ 
\citep{Shapley03, Steidel10}. 

\begin{figure}
\includegraphics[width=0.5\textwidth]{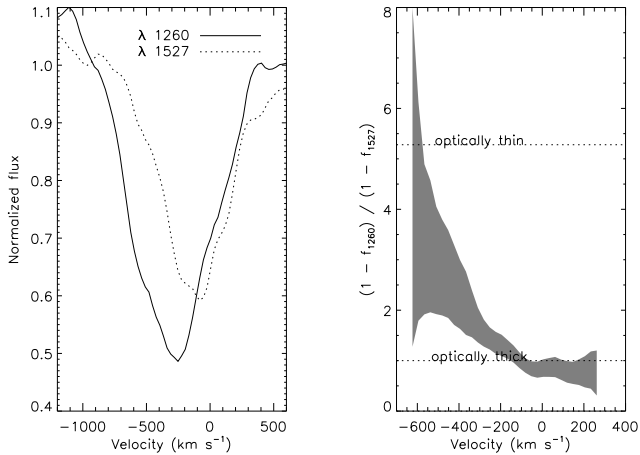}
\caption{\label{fig:SiII_tau} (Left:) comparison of the absorption line profiles for Si{\sc ii} transitions at 1260 and 1527 \AA\ in our composite spectrum. 
(Right:) The profile ratio $(1 - f_{1260})/(1 - f_{1527})$ with the optically thick and thin regimes indicated; the shading refers to the 1$\sigma$ 
uncertainty. Although both transitions are optically thick at low velocities $|v| < 200$ \kms , there is some evidence of optically thin gas at $|v| > 200$ \kms.}
\end{figure}

Of particular diagnostic value are the two relatively unblended transitions of Si{\sc ii} at 1260 and 1527 \AA\ 
whose equivalent width ratio is a valuable tracer of the optical depth \citep{Shapley03}. The profiles of both lines and their ratio are shown 
in Figure~\ref{fig:SiII_tau}. Gas near the systemic velocity ($|v| < 200$ \kms) is clearly optically thick, while for $v<-200$ \kms\, it is intermediate 
in optical depth. This difference is also seen in high-resolution spectra of lensed galaxies \citep{Pettini02, Quider10}, and optically thin gas is seen at large galactocentric radii \citep{Steidel10} suggesting that smaller, optically thin clouds are more easily accelerated to high velocity and large distances. We note that the column density at which both Si{\sc ii} transitions become optically thick ($\tau > 1$) is $N_{\text{Si{\sc ii}}} = 1.9 \times 10^{12}$ cm$^{-2}$. 
Assuming $\log(\text{Si{\sc ii}/H}) = -4.86$ as measured for the lensed galaxy 
cB58 \citep{Pettini02}, we estimate that optically thin absorption occurs in clouds with hydrogen column densities $N_{\text{H{\sc i}}} \lsim 10^{17}$ cm$^{-2}$.

\subsubsection{Fine structure transitions}\label{sec:fs}

A satisfying aspect of our composite spectrum is the successful identification of fine structure emission lines of Si{\sc ii} (see marked features in 
Figure~\ref{fig:composite}). Si{\sc ii} ions in the CGM absorb photons in the resonance transitions and immediately re-emit a photon at 
approximately the same velocity. A photon absorbed at 1260 \AA\, will be re-emitted at either the same wavelength or to the fine structure 
transition Si{\sc ii}*$\lambda$1265. Likewise a photon absorbed at 1527 \AA\, will be re-emitted as either Si{\sc ii}$\lambda$1527 or 
Si{\sc ii}*$\lambda$1533, and an absorbed Si{\sc ii}$\lambda$1304 photon can be re-emitted as Si{\sc ii}*$\lambda$1309. In all three cases 
the probability of emission in the resonant and fine structure transitions is approximately equal. 

Since no absorption is seen in the fine structure transitions, we infer that atoms in the excited ground state will typically decay to the ground state 
before absorbing another photon. Since every absorbed photon is re-emitted, the net equivalent width of the resonant and fine structure transitions 
is $W_{\text{Si{\sc ii}}} + W_{\text{Si{\sc ii}*}} = 0$. The precise equivalent widths depend on the initial absorption $W_{\text{Si{\sc ii},abs}}$ and 
the optical depth. If the gas is optically thin, re-emitted photons will immediately escape giving $W_{\text{Si{\sc ii},em}} = W_{\text{Si{\sc ii}*,em}} 
= - 0.5 W_{\text{Si{\sc ii},abs}}$. In the limit of optically thick gas, resonant photons will be continuously scattered until they emerge as Si{\sc ii}* 
(after 2 scatterings on average). In this case, $W_{\text{Si{\sc ii},em}} = 0$ and $W_{\text{Si{\sc ii}*,em}} = - W_{\text{Si{\sc ii},abs}}$. Since the 
majority of absorption occurs in optically thick gas (\S\ref{sec:lis}), we expect the equivalent width of Si{\sc ii} absorption lines to reflect the kinematics 
and covering fraction of neutral gas with minimal contamination from Si{\sc ii} re-emission.

\subsubsection{The spatial extent of low-ionization absorption}\label{sec:fs_extent}

From the picture above (\S\ref{sec:fs}), we expect that the equivalent width of the fine structure emission lines should be equal and opposite 
to the resonant line equivalent width. This is not the case: $W_{1265} / W_{1260} = -0.56 \pm 0.36$ and $W_{1533} / W_{1527} = -0.52 \pm 0.20$. 
($W_{1309}/W_{1304}$ is more difficult to quantify since Si{\sc ii}$\lambda$1304 is blended with O{\sc i}$\lambda$1302). There are two possible 
explanations. One is that scattered photons have larger path lengths and are subject to greater dust attenuation. Since the UV continuum slopes imply little 
differential extinction (mean E(B-V) $= 0.10$) and we expect only 2 scatterings for the average Si{\sc ii}* photon, this seems unlikely. More reasonably,
the emitting region could be larger than that sampled by our slits. Although the $1\farcs0 \times 1\farcs0$ slit aperture samples most of the continuum light 
and its line of sight absorption, the CGM which absorbs and isotropically re-emits scattered photons extends to much larger radii \citep{Steidel10, Steidel11}. 
The strength of fine structure emission provides a direct measurement of the amount of absorption at small radii contained within the slit. 
The fine structure to resonant absorption line ratio suggests that a fraction $0.53 \pm 0.17$ (combining both $\lambda$1265 and 
$\lambda$1527 measures) of the total fine structure emission is contained within the extraction aperture of 1 arcsec$^2$. The half-light radius 
of Si{\sc ii}* emission, and hence Si{\sc ii} absorption, thus corresponds to $\sim 0\farcs5 = 3.5$ kpc at $z=4$. This scale is only slightly larger than the 
median half-light radius $r_h = 0\farcs31$ of galaxies in our sample, and would be reached in only 20 Myr at the typical outflow velocity (190 \kms). 

It is instructive to reconsider the $z=3$ composite spectrum from \cite{Shapley03} where the fine structure
lines are clearly seen (but were not interpreted fully along the discussion above at the time). The Si{\sc ii}* lines are noticeably stronger in 
our $z\simeq4$ composite, with an average $W_{\text{Si{\sc ii}*}} = 0.7 \pm 0.2$ at $z\simeq4$ c.f. $0.30 \pm 0.05$ \AA\, at $z=3$. Furthermore,
the Si{\sc ii}  absorption lines are {\em weaker} in the $z=4$ composite, suggesting the absorption takes place at larger radii at $z=3$. 
This difference cannot be explained through instrumental differences between the two surveys. Smoothing our $z=4$ composite to match the 3.25 \AA\, resolution of \cite{Shapley03} reduces the equivalent widths by $< 5$\%. Likewise the wider slits used for the $z=3$ data (1\farcs4, 11 kpc) 
c.f. the $z=4$ data (1\farcs0; 7 kpc) is not the cause. If the CGM properties are similar, we expect the fine structure strength to constitute 
a {\em larger} fraction of the resonant absorption line strength in the $z=3$ data, contrary to observations. Defining $R_{FS} = - (W_{1265} + W_{1533}) / (W_{1260} + W_{1527})$, we find $R_{FS} = 0.53 \pm 0.17$ for the $z=4$ composite spectrum and $R_{FS} = 0.16 \pm 0.04$ for $z=3$, indicating a smaller characteristic radius of fine structure emission at $z=4$.

Inescapably, therefore, we conclude the circumgalactic gas around LBGs at $z=4$ differs in two important ways from that at $z=3$: there is 
{\em greater low-ionization absorption} at small radii at $z=4$ and {\em less total low-ionization absorption}. In addressing the origin
of this effect, Figure~\ref{fig:dv_lya_all} shows that the difference in equivalent width is at least partially linked to the kinematic offset from
the \Lya emission. Thus it seems a higher covering fraction at small radii is required to produce the stronger fine structure emission. Higher 
resolution observations of individual bright, or perhaps gravitationally-lensed, galaxies at $z\simeq$4 will ultimately be required to 
separate the relative contributions of covering fraction and kinematics in explaining this result.

\subsection{High-ionization lines}

The high-ionization lines Si{\sc iv} and C{\sc iv} arise both in interstellar gas and in stellar P-Cygni winds. The velocity centroid of 
Si{\sc iv}, $v = 140$ \kms, is consistent with the low-ionization interstellar absorption lines suggesting that most absorption is interstellar 
in origin. This is supported by the absence of a significant redshifted emission component expected for a P-Cygni profile. In 
contrast, C{\sc iv} is broader with a larger absorption velocity offset $v = -370$ \kms and redshifted P-Cygni emission indicating a large contribution from stellar winds. We also detect the N{\sc v}$\lambda\lambda$1239,1243 P-Cygni feature, although the proximity to \Lya makes this feature difficult to study in detail.

We can determine the optical depth of highly ionized outflowing gas from the ratio of Si{\sc iv} absorption lines. The ratio $W_{\text{Si{\sc iv}}\lambda1394} / W_{\text{Si{\sc iv}}\lambda1403} = 2.0$ for optically thin absorption, and 1.0 in the optically thick case. \cite{Shapley03} find optically thin absorption in composite spectra of LBGs at $z=3$, whereas we measure a ratio $1.4 \pm 0.4$ in the composite spectrum shown in Figure~\ref{fig:composite} indicating a significant contribution of optically thick absorption at $z\sim4$. The total equivalent width of the Si{\sc iv} doublet is weaker by a factor $0.72 \pm 0.12$ in Figure~\ref{fig:composite} compared to the $z=3$ composite of \cite{Shapley03}. The combination of higher optical depth and lower equivalent width suggests that the velocity range and/or covering fraction of the ionized gas traced by Si{\sc iv} is lower at higher redshift.

\subsubsection{Metallicity}\label{sec:metallicity}

\begin{figure}
\includegraphics[width=0.5\textwidth]{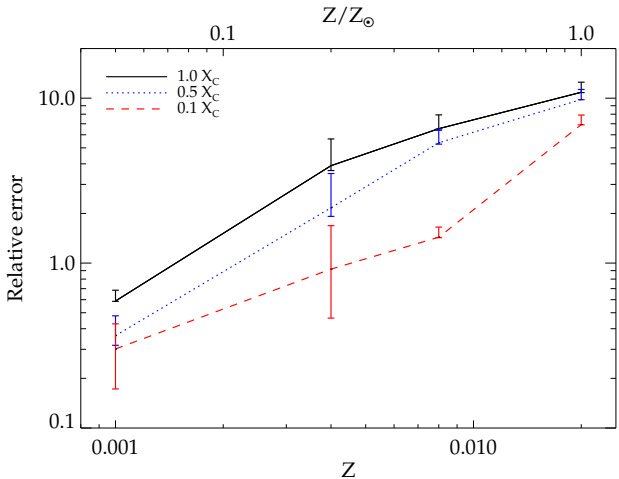}
\caption{\label{fig:bpass} Relative error in the equivalent widths of C{\sc iv} and He{\sc ii} predicted by the stellar population synthesis code {\sc BPASS} and values measured for the composite spectrum shown in Figure~\ref{fig:composite} (see text for details). Relative error values $\lsim 1$ indicate good agreement. Model results for a range of metallicity $Z$ and carbon depletion factor $X_C$ as a function of age are described in \cite{Eldridge11}. 
The values $X_C = 0.5$ (0.1) correspond to a carbon abundance reduced by a factor of 2 (10) relative to solar abundance ratios.
Lines are plotted for the median age of our sample ($10^{8.5\pm0.6}$ years) determined from spectral energy density models with the same initial mass function and star formation history used in {\sc BPASS}, and error bars indicate the allowed range for ages within $1\sigma$ of the median.
Measurements of C/O abundance at both low and high redshift indicate carbon depletion factors $X_C \simeq 0.2-0.5$ in galaxies with $Z<0.2\, Z_{\odot}$ (e.g. \citealt{Erb10}).
For this range of $X_C$, models with $Z \lsim 0.004$ (equivalent to $Z = 0.2\, Z_{\odot}$) are in good agreement with the data. Solar metallicity models are unable to reproduce the observed equivalent widths.}
\end{figure}

The P-Cygni profile of C{\sc iv} is sensitive to metallicity, and the combination of C{\sc iv} and He{\sc ii} equivalent widths constrains both the age and metallicity. Here we compare the equivalent widths measured in the composite spectrum (Figure~\ref{fig:composite}) with theoretical models in order to estimate the typical metallicity of galaxies in our sample. We note that the equivalent width of C{\sc iv} contains significant interstellar absorption, so the value reported in Table~\ref{tab:lines} should be treated as an upper limit on the P-Cygni component. Assuming the interstellar absorption component of C{\sc iv} is similar to that of Si{\sc iv} ($W\simeq1.0$ \AA), we take the P-Cygni absorption component to have equivalent width $-1.6 \pm 1.0$ \AA\, with a conservative uncertainty. We compare this estimate and the measured equivalent width of He{\sc ii} (reported in Table~\ref{tab:lines}) with predictions from the stellar population synthesis code {\sc BPASS} presented in \cite{Eldridge09}. 
We consider {\sc BPASS} models which include binary evolution with continuous star formation rate, and determine the difference $\Delta W$ between observed and predicted equivalent width as a function of metallicity and relative carbon abundance. We restrict the stellar population age to be within the 1$\sigma$ scatter of the median value for galaxies in our sample, determined to be $10^{8.5 \pm 0.6}$ years from spectral energy density fits assuming constant star formation with a \cite{Kroupa02} initial mass function.
Figure~\ref{fig:bpass} shows the resulting relative error between {\sc BPASS} models and measured equivalent width, which we define as 
\begin{equation}
\text{Relative error}
= 0.5 \left[ \left(\frac{\Delta W_{\text{C{\sc iv}}}}{\sigma_{\text{C{\sc iv}}}} \right)^2 + \left(\frac{\Delta W_{\text{He{\sc ii}}}}{\sigma_{\text{He{\sc ii}}}} \right)^2 \right].
\end{equation}
Relative error values $\lsim 1$ are thus consistent with the data.

Models with $Z \leq 0.004$ and somewhat depleted carbon abundance are in good agreement with the data. We note that observations of both local and high-redshift galaxies indicate typical relative carbon abundances (C/O) $\simeq 2-5\times$ lower than the solar value (equivalent to $X_C = 0.2-0.5$ in Figure~\ref{fig:bpass}) for metallicities $Z \lsim 0.004$ \citep{Kobulnicky98, Shapley03, Erb10, Eldridge11}. Solar metallicity models ($Z = 0.020$) do not fit the data. We conclude that the typical metallicity of galaxies in our sample is $Z \lsim 0.004$ or $\lsim 0.2 \times$ solar metallicity. Bright galaxies at $z<3.8$ in the FORS2 sample have gas-phase metallicity $12+\log(O/H) = 7.7-8.5$ or about $0.1-0.6 \times$ the solar value, in reasonable agreement \citep{Maiolino08}. These values are consistent with the allowed metallicity range $0.1 < Z/Z_{\odot} < 0.6$ inferred for the \cite{Shapley03} composite using the same {\sc BPASS} models \citep{Eldridge11}. Given that the equivalent widths of C{\sc iv} and He{\sc ii} in \cite{Shapley03} agree well with those in Table~\ref{tab:lines}, we expect the typical metallicity of galaxies to be similar in both samples.

\section{Spectroscopic trends}\label{sec:trends}

 We now turn to an analysis of how the various spectroscopic features discussed in \S4 are related to observable properties of LBGs
as a prelude to considering how they might evolve with redshift. We will begin with the dependence of low-ionization absorption 
line strength with \Lya equivalent width. This is motivated by the common physical dependence of these features -- both are governed by the kinematics and covering fraction of neutral circumgalactic gas -- and also because this has been examined in detail for LBGs at $z=3$ \citep{Shapley03}.

\begin{figure}
\includegraphics[width=0.5\textwidth]{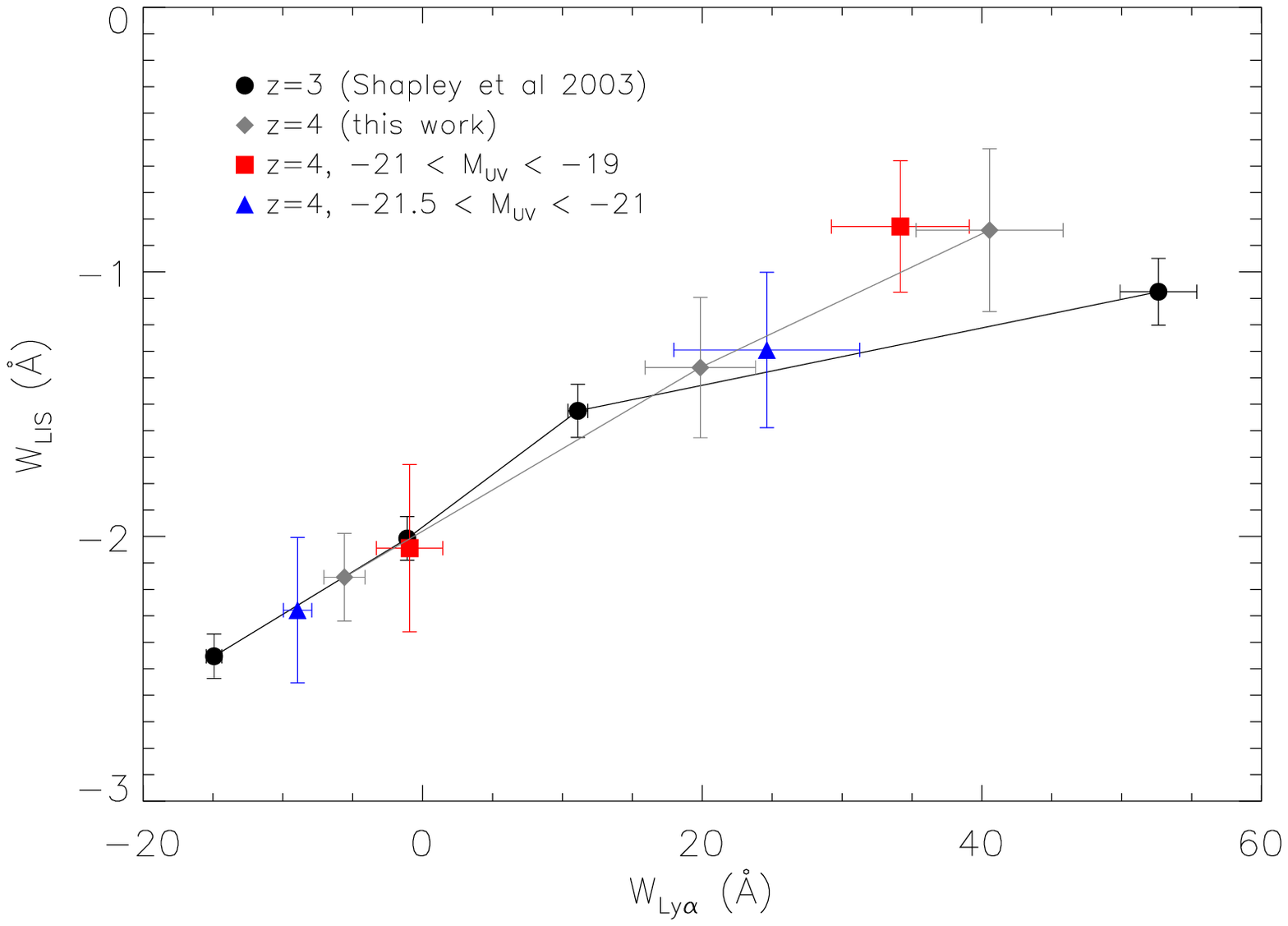}
\caption{\label{fig:lis_lya} \label{fig:lis_lya_muv} 
Equivalent width of low-ionization absorption lines compared to that of \Lya. Gray diamonds are from our sample at $z\simeq4$ divided into bins of $W_{\Lya} < 0$, $W_{\Lya} = 0-40$ \AA, and $W_{\Lya} > 40$ \AA. Black circles show the sample of Shapley et al (2003) at mean $z=3$, divided into quartiles of $W_{\Lya}$. Galaxies in the $z=3$ sample have typical luminosities $-21.5 < M_{UV} < -21.0$ corresponding to the blue triangles. Galaxies of the same luminosity lie on the same correlation between $W_{\Lya}$ and $W_{LIS}$ at both $z=3.9$ and $z=3$. Fainter galaxies have weaker low-ionization absorption lines at fixed $W_{\Lya}$.}
\end{figure}

We construct composite spectra of three sub-samples of galaxies divided according to their \Lya equivalent width. Defining $W_{LIS}$ as the average equivalent width of 
$\lambda$1260, $\lambda$1303, $\lambda$1334, and $\lambda$1527 \AA\ features, Figure~\ref{fig:lis_lya} shows weaker $W_{LIS}$ for galaxies with
stronger \Lya. This was also noted for $z=4$ LBGs by \cite{Vanzella09}.
Both the $z$=3 and $z$=4 samples show this trend, although $W_{LIS}$ is weaker at $z=4$ for galaxies with strong \Lya emission. This difference can 
be attributed to the luminosity-dependent trend of stronger $W_{\Lya}$ and weaker $W_{LIS}$ in fainter galaxies (Paper I; \citealt{Shapley03}; \citealt{Vanzella09}). 
If we consider a subset of galaxies in our sample with similar absolute magnitudes to those observed at $z=3$ ($-21.5 < M_{UV} < -21.0$), 
we recover the same normalization (Figure~\ref{fig:lis_lya_muv}). Our sample is 90\% complete at the corresponding apparent magnitudes, 
so we expect a negligible bias. 

\begin{figure}
\includegraphics[width=0.5\textwidth]{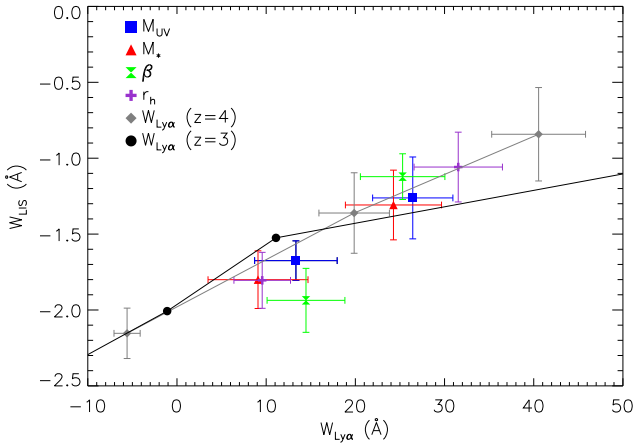}
\caption{\label{fig:lis_lya_all} Equivalent width of low-ionization absorption lines compared to that of \Lya, divided according to observable demographic properties as described in the text. Points binned by $W_{\Lya}$ at $z=3$ and $z=4$ are the same as in Figure~\ref{fig:lis_lya_muv}. The sample was divided into two bins for each demographic property ($M_{UV}$, $M_*$, $\beta$, and $r_h$), with $W_{\Lya}$ and $W_{LIS}$ measured from composite spectra of the galaxies in each bin.}
\end{figure}

Having established the correlation of $W_{LIS}$ with $W_{\Lya}$, we now examine the dependence with other demographic properties. 
We divide the full sample into two bins of equal size according to each property of interest. The results are shown in Figure~\ref{fig:lis_lya_all}. 
We briefly review the trend of each property with $W_{LIS}$ and $W_{\Lya}$ and discuss the physical origin. Many of these trends were
seen in \cite{Shapley03} and \cite{Vanzella09} and are clearly inter-related due to correlations between the demographic properties.

Less luminous galaxies have stronger $W_{\Lya}$ and weaker $W_{LIS}$ illustrating that higher star formation rates drive larger amounts 
of neutral gas into the CGM with higher velocity and/or covering fraction. Defining the ultraviolet spectral slope as $\beta = 5.30 \cdot (i_{AB} - z_{AB}) - 2.04$ for 
B-drops \citep{Bouwens09}, we also find that bluer galaxies with lower $\beta$ have stronger $W_{\Lya}$ and weaker $W_{LIS}$. 
This trend was also noted in Paper I, which showed that LBGs with strong \Lya emission have systematically bluer $\beta$.
Since neutral gas also presumably contains dust, the same gas which gives rise to $W_{LIS}$ also reddens the continuum. Stellar masses $M_*$ are
measured for 60\% of our sample for which there is unconfused {\sl Spitzer}/IRAC photometry. As expected from the trends with luminosity, 
lower mass galaxies have stronger $W_{\Lya}$ and weaker $W_{LIS}$. Finally, measuring half-light radii $r_h$ from the GOODS ACS data 
with {\sc Sextractor}, we find smaller galaxies have stronger $W_{\Lya}$ and weaker $W_{LIS}$. 

Could these trends could be due to selection effects or sample bias? The trend of stronger \Lya for less luminous LBGs 
is of particular concern, since fainter galaxies will require a larger $W_{\Lya}$ for detection. To address this, we consider only the 32 galaxies 
with apparent magnitudes $z'_{AB} < 25$ for which the spectroscopic sample is 90\% complete. This sub-sample is divided into two equal bins and 
the results confirm that trends seen in the larger sample also hold in brighter galaxies unaffected by sample bias.

The trends shown in Figure~\ref{fig:lis_lya_all} are generally consistent with the overall trend of $W_{LIS}$ with $W_{\Lya}$ 
(Figure~\ref{fig:lis_lya}). The $W_{\Lya}$ variations mostly arise from the distribution of neutral gas (traced by $W_{LIS}$), with relatively little 
effect from other demographic properties examined ($M_{UV}$, $\beta$, $M_*$, $r_h$). However, demographic properties do have some effect.
The strongest deviation seen in Figure~\ref{fig:lis_lya_all} is that with $\beta$, which shows stronger $W_{LIS}$ than would be expected at a 
given $W_{\Lya}$ for galaxies with red UV slopes. Noted also by \cite{Shapley03}, this suggests that outflowing neutral gas contains 
dust which reddens the continuum. Also, more luminous galaxies (i.e. those with higher star formation rates) have stronger $W_{LIS}$ at a 
given $W_{\Lya}$ (Figure~\ref{fig:lis_lya_muv}). This is likely due to increased absorption at large velocities in galaxies with higher SFR, 
as observed at $z\simeq 1.4$ \citep{Weiner09}. For all other demographics, the composite spectra have values of $W_{LIS}$ within 
$1\sigma$ of that expected purely based on $W_{\Lya}$. 

In summary, the trends seen in Figure~\ref{fig:lis_lya_all} arise almost entirely because of variations in the neutral gas covering 
fraction and/or kinematics, which are themselves correlated with the demographic properties.

\subsection{Kinematics}

We have now established that outflowing neutral gas is the dominant factor in determining both $W_{\Lya}$ and $W_{LIS}$. However, variations from 
the trend of $W_{LIS}$ with $W_{\Lya}$ are apparent, particularly with $\beta$ and $M_{UV}$ as discussed above. The most obvious mechanism for 
this behaviour is a systematic difference in the covering fraction $f_c$ and kinematics of neutral gas. Assuming $W_{LIS} \propto \int f_c(v) dv$ (where 
$v$ is the outflow velocity), a higher $f_c$ and lower velocity range can conspire to give a constant $W_{LIS}$. However, changing $f_c$ and $v$ will also 
affect the transmission of \Lya photons resulting in a different $W_{\Lya}$. It is therefore of interest to consider how to distinguish between the covering 
fraction and kinematics of neutral gas.

\begin{figure}
\includegraphics[width=0.5\textwidth]{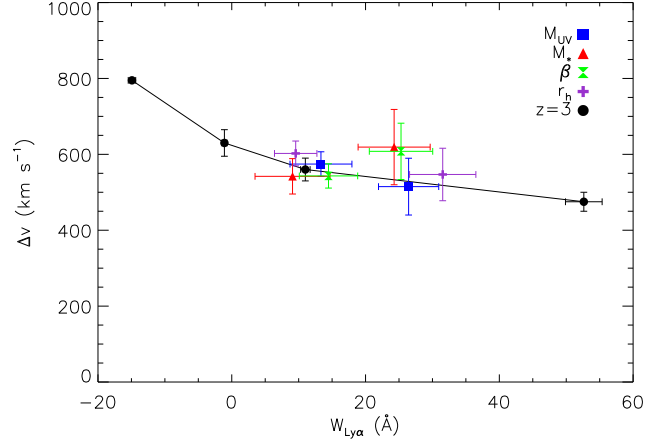}
\caption{\label{fig:dv_lya_all} Kinematic offset $\Delta v$ between \Lya emission and low-ionization absorption lines, as a function of $W_{\Lya}$. The composites used to measure $\Delta v$ and $W_{\Lya}$ are the same as in Figure~\ref{fig:lis_lya_all}. Data at $z=3$ are from \cite{Shapley03}.}
\end{figure}

The kinematics of neutral gas can be roughly parameterized by the velocity dispersion and centroid of low-ionization interstellar absorption lines. We 
have measure the velocity $\Delta v$ of low-ionization absorption lines with respect to \Lya as a proxy for outflow velocity. $\Delta v$ is strongly correlated 
with $W_{\Lya}$ and $W_{LIS}$ in LBGs at $z=3$ \citep{Shapley03} in the sense that larger velocities are associated with stronger interstellar 
absorption and weaker \Lya emission. Figure~\ref{fig:dv_lya_all} shows $\Delta v$ measured from the same composite spectra used to determine
demographic trends (e.g. Figure~\ref{fig:lis_lya_all}), as well as the results from \cite{Shapley03}. All $z\simeq4$ composites are consistent 
(within $1\sigma$) with the relation measured at $z=3$ as well as the mean $\Delta v$ measured for B-dropout galaxies by \cite{Vanzella09}.
We measure a velocity dispersion for each composite, and find that each is within $1.2\sigma$ 
of the effective spectral resolution $\sigma = 290$ \kms (measured for the stellar [C{\sc iii}] feature in Figure~\ref{fig:composite}). We are therefore
unable to detect trends in outflow kinematics with demographic properties in composite spectra. Higher signal-to-noise data, higher spectral resolution, 
or detailed studies of individual galaxies are required to address trends in kinematics and covering fraction of neutral gas at $z=4$.

\section{The evolving CGM}\label{sec:cgm}

We now turn to a discussion of the redshift evolution of the neutral CGM surrounding typical LBGs. The most useful probes are \Lya and low-ionization absorption lines which trace the kinematics and covering fraction of neutral gas, and fine structure emission lines which provide a constraint on the spatial extent of the absorbing gas (\S\ref{sec:fs_extent}).

The common dependence of \Lya and low-ionization absorption lines on the neutral CGM results in a strong correlation between $W_{LIS}$ and $W_{\Lya}$. Various physical properties of LBGs are correlated with both $W_{LIS}$ and $W_{\Lya}$, but in such a way that the relation between $W_{LIS}$ and $W_{\Lya}$ remains nearly constant (Figure~\ref{fig:lis_lya_all}). Furthermore, the $W_{LIS}$--$W_{\Lya}$ relation at fixed $M_{UV}$ does not change significantly with redshift between $z=3$ and $z=4$ (Figure~\ref{fig:lis_lya_muv}).

\begin{figure}
\includegraphics[width=0.5\textwidth]{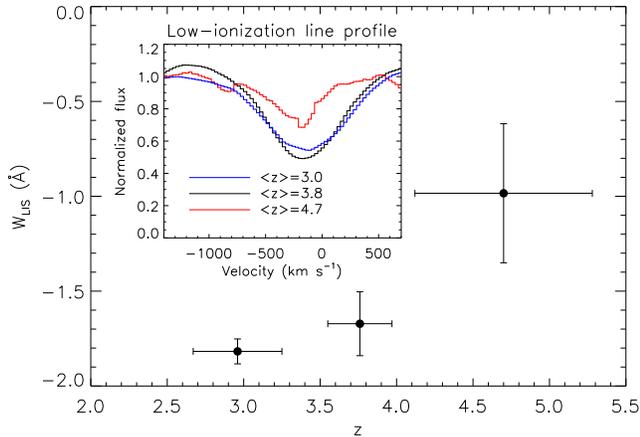}
\caption{\label{fig:lis_z} Equivalent width of low-ionization absorption lines measured from composite spectra of LBGs at different redshifts. The DEIMOS and FORS2 data presented in this paper are separated into two sub-samples of equal size as described in the text. We also show the equivalent result at $z=3$ from the composite spectrum of \cite{Shapley03}, with redshift distribution described in \cite{Steidel03}. The average low-ionization line profile of each composite is shown in the inset. Low-ionization absorption lines are significantly weaker for galaxies in the highest redshift composite. All three samples have consistent mean luminosity and $W_{\Lya}$.}
\end{figure}

To further examine evolutionary trends with redshift, we now divide our spectroscopic sample into two bins of redshift at fixed $M_{UV}$, now including galaxies at all redshifts (no longer restricted to $z<4.5$ as in previous sections). We consider galaxies with absolute magnitude $-21.5<M_{UV}<-20.5$, chosen to be representative of the sample in \cite{Shapley03}. There are 64 galaxies in our sample within this $M_{UV}$ range (see Figure~\ref{fig:sample}). We construct composite spectra of galaxies with redshift above and below the median $z=4.1$ and measure the equivalent width of $\Lya$ and low-ionization lines (both resonant absorption and fine structure emission) in each composite. The results are given in Table~\ref{tab:redshift} along with the demographic properties of galaxies in each sub-sample. The quantities $M_{UV}$, $M_*$, and $r_h$ are measured from photometry while $\beta$ is determined from a direct fit to the ultraviolet continuum in the composite spectrum. We define $f_{\lambda} \propto \lambda^{\beta}$ and fit the rest frame $1300-1700$ \AA\ to determine $\beta$, with uncertainty quantified using the bootstrap method described in \S\ref{sec:composite}. Aside from redshift, the demographics of each sub-sample are quite similar. The higher redshift galaxies have slightly higher average $M_*$, smaller $r_h$, and smaller (bluer) $\beta$. $W_{\Lya}$ is consistent for both to within the sample variance, and is also consistent with the value $W_{\Lya} = 14.3$ \AA\, measured for the composite spectrum of $z=3$ LBGs in Shapley et al (2003). 
The most striking difference is in the strength of the low-ionization absorption lines, which are significantly weaker at higher redshifts (Figure~\ref{fig:lis_z}). The variation in $W_{LIS}$ is {\em not} explained by systematic differences in $W_{\Lya}$ or demographic properties, hence we seek an alternate explanation.

We first examine whether the evolution in $W_{LIS}$ could arise as a result
of different equivalent width distributions for \Lya. Although the mean
$W_{\Lya}$ across our two redshift subsamples is similar, the lower
redshift subsample has a broader distribution and contains
more galaxies with \Lya in absorption ($W_{\Lya}<0$). This is reflected in the larger sample variance in $W_{\Lya}$ at lower redshift (Table~\ref{tab:redshift}). 
We evaluate the effect of this potential bias on $W_{LIS}$ by constructing 
a composite spectrum from a subset of the $z<4.1$ galaxies with intermediate $W_{\Lya} = 0-30$ \AA, resulting in a consistent mean $W_{\Lya}$ with sample variance reduced by a factor of 2.5. This composite has $W_{LIS}=-1.5$ \AA, 0.2 \AA\ higher than when the full range of $W_{\Lya}$ is used, but still considerably lower than the value $W_{LIS}=-1.0$ \AA\ measured for the higher redshift galaxies. 
We therefore conclude that differences in the $W_{\Lya}$ 
distribution are insufficient to explain
the observed variation in absorption line strength with redshift.

There are several possible physical explanations for the evolution of $W_{LIS}$ with redshift shown in Figure~\ref{fig:lis_z}. 
One possibility is that the kinematics and/or covering fraction of neutral gas are systematically different. For example, an outflowing wind with fixed input energy and momentum will reach higher velocity at lower redshifts due to lower density of the IGM. However, this effect should be stronger between $z=3.0-3.8$ than from $z=3.8-4.7$, whereas the decrement in $W_{LIS}$ is much stronger from $z=3.8-4.7$ (Figure~\ref{fig:lis_z}). Furthermore, we measure a higher offset between the velocity centroid of \Lya emission and low-ionization absorption for $z>4.1$ galaxies ($\Delta v = 660 \pm 80$) than for $z<4.1$ ($\Delta v = 550 \pm 40$). Both measurements are consistent with the trend shown in Figure~\ref{fig:dv_lya_all}. Based on the trend of $W_{LIS}$ with $\Delta v$ seen at $z=3$ \citep{Shapley03}, we would then expect {\em stronger} $W_{LIS}$ at higher redshift. Kinematics are thus unable to explain the difference in $W_{LIS}$, at least with the information currently available. Data with higher spectral resolution and signal-to-noise are needed to fully address differences in the covering fraction and kinematics of neutral gas.

\begin{figure}
\includegraphics[width=0.5\textwidth]{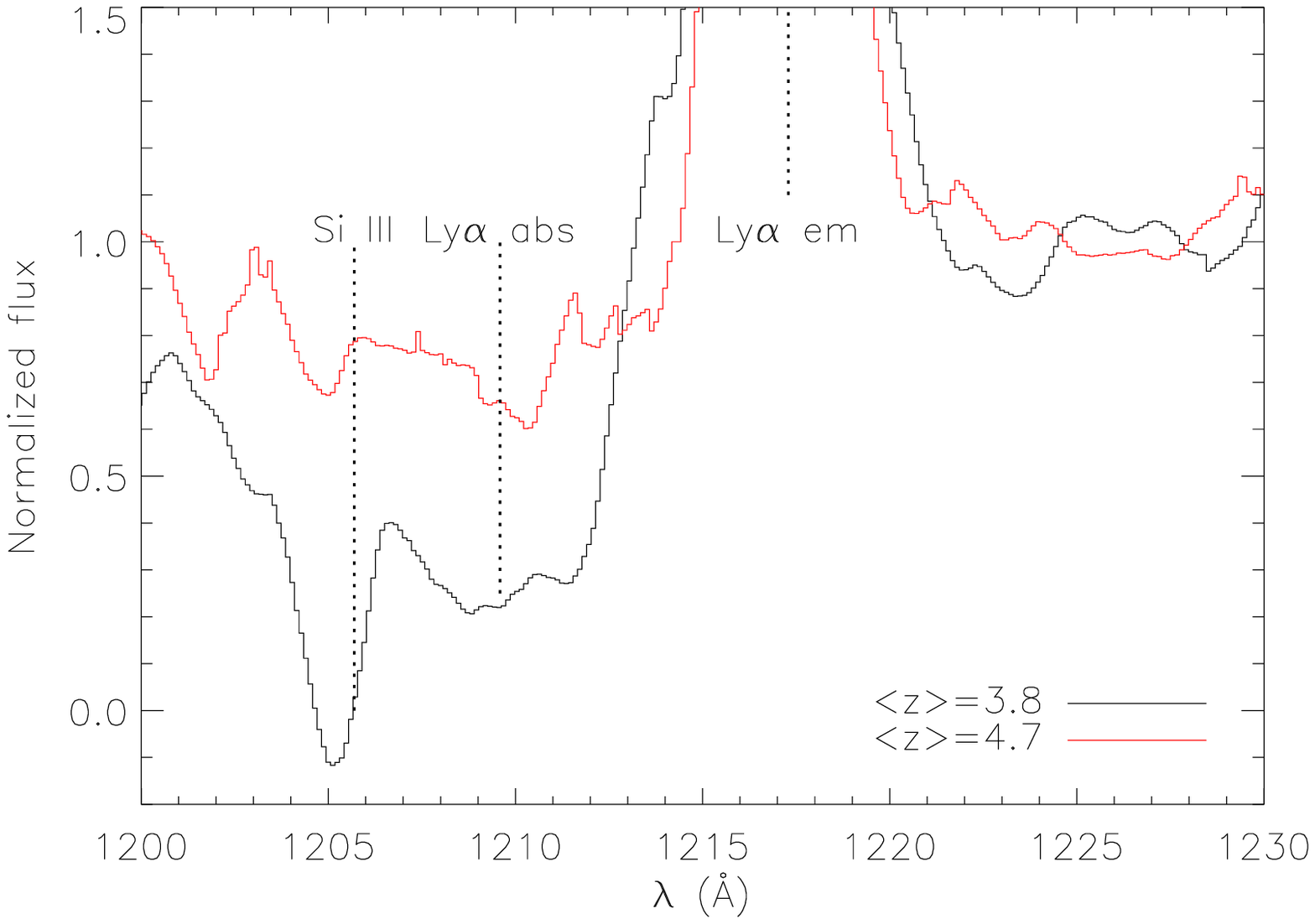}
\caption{\label{fig:lya_z} Composite spectra of LBGs showing the region around \Lya for the two redshift ranges described in \S\ref{sec:cgm}. We show the position of Si{\sc iii}$\lambda1206$ blueshifted by 200 \kms, \Lya emission redshifted by 400 \kms, and \Lya absorption blueshifted by 1500 \kms\, where it is seen most prominently. The \Lya absorption trough is weaker in the higher redshift composite indicating a lower incidence of neutral gas with high column density. Si{\sc iii} absorption is also weaker in the higher redshift composite.}
\end{figure}

Another obvious potential cause of weaker $W_{LIS}$, or equivalently decreased $W_{\Lya}$ at a fixed $W_{LIS}$, is the presence of neutral hydrogen with very low column density of heavy elements. In this case we would expect optically thin absorption line profiles. This could be directly tested as shown in \S\ref{sec:fs}, but the signal-to-noise of Si{\sc ii}$\lambda1527$ in the high redshift composite is too low to constrain the optical depth. 
We do see some evidence that the column density of neutral gas is lower at higher redshifts based on \Lya line profiles in the composite spectra, shown in Figure~\ref{fig:lya_z}. The \Lya absorption trough at $\sim1200-1215$ \AA\, is significantly weaker in the higher redshift composite. This absorption arises at least in part from damping wings of high column density gas with $N_{\text{H{\sc i}}} \gsim 10^{20}$ cm$^{-2}$ associated with the low-ionization metal absorption lines (e.g. \citealt{Pettini00,Pettini02}). The higher redshift galaxies are therefore characterized by lower typical $N_{\text{H{\sc i}}}$ or/and a lower covering factor of high column density gas. Higher signal-to-noise data at $z\simeq5$ is required to determine whether this affects the optical depth of low-ionization metal transitions.

Weaker low-ionization lines could also be caused by a systematically higher ionization state at higher redshifts resulting in lower column densities of low-ionization gas. This would also explain the weaker damped \Lya absorption trough seen at higher redshifts (Figure~\ref{fig:lya_z}). This scenario can be tested by measuring the equivalent widths of higher-ionization silicon transitions, in particular Si{\sc iii}$\lambda$1206 and Si{\sc iv}$\lambda\lambda$1393,1402. We find that the Si{\sc iv} lines are weaker at higher redshift by a factor of $0.68 \pm 0.25$ in $W_{\text{Si{\sc iv}}\lambda\lambda1393,1402}$, consistent with the decrement in $W_{\text{Si{\sc ii}}}$. We do not detect the Si{\sc iii}$\lambda$1206 line in the high redshift composite, with a 1$\sigma$ upper limit of $W_{\text{Si{\sc iii}}} < 0.8$ times that of the lower redshift composite. Si{\sc iii}$\lambda$1206 is marked in Figure~\ref{fig:lya_z} and is clearly stronger at lower redshift. We therefore find no evidence of a significant change in the ionization state of outflowing gas at higher redshifts.

A final possibility for lower column density of heavy elements is that \Lya is scattered by ``cold streams'' of nearly metal-free neutral gas, predicted by simulations to accrete onto galaxies at small radii and with increasing rates at higher redshift (e.g. \citealt{Dekel09, Faucher11}). If they exist, such streams could be identified as H{\sc i} absorption systems with no corresponding absorption from heavy elements. 
\cite{Steidel10} find little evidence for the presence of such streams around LBGs at $z=2-3$. A similar study at higher redshift would be very interesting but also extremely challenging with current observational facilities.

Could the difference in $W_{LIS}$ with redshift be caused in part by a different physical extent of circumgalactic gas? In \S\ref{sec:fs_extent} we showed that fine structure emission lines can be used to measure the amount of absorption taking place within the size of the spectroscopic aperture. This quantity does appear to change with redshift, in the sense that absorption takes place at smaller characteristic radius in LBGs at higher redshift. The composite spectra of LBGs at $z=4$ has weaker absorption lines and stronger fine structure emission than the $z=3$ composite of \cite{Shapley03} (Figure~\ref{fig:composite2}), in qualitative agreement with trends seen at $z=3$ (\citealt{Shapley03}; see their Figure 9). We have only a weak constraint on the redshift evolution at $z>4$ due to limited signal-to-noise in the high redshift composite. The ratio of total Si{\sc ii} fine structure emission $W_{\text{Si{\sc ii}}^*}$ in the high- and low-redshift composites is $1.1 \pm 0.5$ (taken as
\begin{eqnarray*}
\frac{W_{1265}+W_{1309}+W_{1533}}{W_{1260}+W_{1303}+W_{1527}}
\end{eqnarray*}
with values given in Table~\ref{tab:redshift}), while the corresponding ratio of absorption line strength $W_{LIS}$ is $0.6 \pm 0.2$. The inferred fraction of low-ionization absorption within the slit aperture is thus a factor $1.8 \pm 1.0$ larger in the higher redshift sample, consistent with no evolution. Higher signal-to-noise is required to accurately constrain the spatial extent of low-ionization absorption at $z>4$.

To summarize, we find weaker low-ionization absorption lines in LBGs at higher redshift with fixed $M_{UV}$. The difference in $W_{LIS}$ is not consistent with the trends observed for \Lya and various demographic properties examined in \S\ref{sec:trends}. We have discussed several possible causes for this discrepancy including variation in the kinematics, covering fraction, optical depth, ionization state, and spatial extent of absorbing gas. The data show that the ionization state has no significant effect, but we are unable to conclusively address other possible causes. Spectra of galaxies at $z>4$ with higher signal-to-noise are required to more accurately constrain the optical depth and spatial extent. 
Additionally, spectra of individual galaxies taken 
with higher spectral resolution will be required to independently
determine the covering fraction and kinematic structure of
absorbing gas.

\subsection{Galaxy evolution}

We argue in \S\ref{sec:trends} that \Lya equivalent width is determined primarily by the neutral CGM, which is correlated with various demographic galaxy properties (Figure~\ref{fig:lis_lya_all}). It is well established that these demographic properties vary with redshift and we expect $W_{\Lya}$ to vary accordingly. Large photometric surveys have shown that Lyman break galaxies at increasingly higher redshifts $z>3$ have lower luminosities \citep{Bouwens07,Bouwens11a}, bluer UV spectral slopes \citep{Bouwens09, Bouwens11b}, smaller stellar masses \citep{Stark09, Gonzalez11}, and smaller sizes \citep{Bouwens04, Ferguson04}. This is in accordance with inside-out galaxy growth: galaxies increase in size and stellar mass as they evolve with time, while increasing metallicity and dust content reddens the ultraviolet continuum. Simultaneously, star formation drives large-scale outflows of gas which reach larger distances and are accelerated to larger velocities at later times (e.g. \citealt{Murray10}). Galaxies which are more evolved (i.e. larger, more massive, redder) should therefore have a CGM characterized by larger spatial extent, larger velocity range, and higher covering fraction of neutral gas. Observationally this results in weaker $W_{\Lya}$, stronger $W_{LIS}$, and relatively weaker fine structure emission. These are precisely the trends observed at both $z=3$ \citep{Shapley03} and $z=4$ (this work).

\subsection{\Lya in the epoch of reionization}

We reiterate that Lyman break galaxies at increasingly higher redshifts $z>3$ have lower luminosities, bluer UV spectral slopes, smaller stellar masses, and smaller sizes. Notably, {\em all trends in the demographics of galaxies at higher redshift are correlated with stronger $W_{\Lya}$ and weaker $W_{LIS}$} (Figure~\ref{fig:lis_lya_all}). We therefore expect typical galaxies at higher redshifts to have, on average, stronger \Lya emission. Earlier results from this survey confirm that strong \Lya emission is more frequent in galaxies at higher redshift (Paper I; Paper II). We find no evidence of this trend reversing. In fact, galaxies with extremely small size, low mass, and blue $\beta$ tend to be the strongest \Lya emitters (e.g. \citealt{Erb10}). We do, however, expect the average \Lya emission strength to decrease significantly at increasing redshifts in the epoch of reionization due to neutral hydrogen in the IGM (e.g. \citealt{Haiman99}).

We have devoted considerable discussion to the properties of \Lya in part because \Lya is of great interest as a tracer of cosmic reionization. Several authors have now presented evidence that reionization was incomplete at $z\simeq7$ based on a rapidly decreasing fraction of galaxies with strong \Lya emission at $z\gsim6.5$ \citep{Schenker11, Ono11, Pentericci11}. Although trends at $z=3-4$ suggest that the galaxies observed by these authors should have a higher fraction of strong \Lya emission, we have presented evidence that the relation between $W_{\Lya}$ and $W_{LIS}$ is systematically different at $z>4$ (Figure~\ref{fig:lis_z}). This implies a systematic difference in the spatial, kinematic, or optical depth structure of neutral circumgalactic gas compared to galaxies at lower redshift. The physical origin of this evolution and its effect on $W_{\Lya}$ at $z>4$ will need to be understood in order to fully interpret the results of \Lya surveys at higher redshifts in the context of reionization.

\section{Summary}

The rest-frame ultraviolet spectra of star-forming galaxies contains a wealth of information about the properties of the circumgalactic medium. In this paper we have presented an analysis of several features which trace the CGM with a focus on the properties of neutral gas. We find that the trends observed at lower redshift ($z=3$; \citealt{Shapley03}) also hold at $z=4$ with approximately the same normalization (Figure~\ref{fig:lis_lya_muv}). However, we find evidence for rapid evolution at $z>4$ with lower $W_{LIS}$ at fixed $W_{\Lya}$ and $M_{UV}$, suggesting a systematic difference in the spatial distribution, kinematics, ionization state, or optical depth of circumgalactic gas at higher redshifts. 
We determine that the ionization state is not responsible
for the observed evolution but are unable to distinguish between
kinematics, covering fraction, optical depth, or the spatial extent
of neutral gas as the likely cause.
We are collecting additional spectra of LBGs at $z>4$ with our ongoing survey, including high spectral resolution observations of bright lensed galaxies from which we can disentangle the kinematic profile and covering fraction of neutral gas. These data will allow us to address the precise magnitude and physical origin of evolution in circumgalactic gas properties.

As a final note, we emphasize that the neutral CGM is of great interest in the context of reionization of the universe. Neutral gas in the circumgalactic medium absorbs ionizing radiation, thereby inhibiting the ability of galaxies to reionize the universe. The escape fraction of ionizing photons is one of the most important and uncertain factors in determining the contribution of star-forming galaxies to reionization \citep{Robertson10}. We have shown that typical galaxies at higher redshift have weaker low-ionization absorption lines based on their demographic trends, and presented new evidence that absorption lines are systematically weaker at $z>4$ even for fixed demographic properties. This is likely caused by a lower covering fraction and/or velocity range of neutral gas, and we will address the physical origin of this evolution with future data from our ongoing survey. Determining the redshift evolution of neutral gas covering fraction in LBGs will be of great interest for interpreting surveys of \Lya emission in the context of reionization and addressing the role of star-forming galaxies in reionizing the universe.

\section*{Acknowledgments}
D.P.S. acknowledges support from NASA through Hubble Fellowship grant \#HST-HF-51299.01 awarded by the Space Telescope Science Institute, which is operated by the Association of Universities for Research in Astronomy, Inc, for NASA under contract NAS5-265555. 
R.S.E. acknowledges the hospitality of Piero Madau and colleagues at the University of California, Santa Cruz where this work was completed. 
We thank Masami Ouchi, Chuck Steidel, Max Pettini, Anna Quider, Crystal Martin, Alice Shapley, and Gwen Rudie for helpful discussions.
The analysis pipeline used to reduce the DEIMOS data was developed at UC Berkeley with support from NSF grant AST-0071048.
Most of the data presented herein were obtained at the W.M. Keck Observatory, which is operated as a scientific partnership among the California Institute of Technology, the University of California and the National Aeronautics and Space Administration. The Observatory was made possible by the generous financial support of the W.M. Keck Foundation.
The authors wish to recognize and acknowledge the very significant cultural role and reverence that the summit of Mauna Kea has always had within the indigenous Hawaiian community.  We are most fortunate to have the opportunity to conduct observations from this mountain.

\begin{table}[p]
\begin{tabular}{lccc}
\hline
\hline
Ion  &  $\lambda_{\text{rest}}$  &  $W$  &  $v_{\text{cen}}$ \\
     &  (\AA)                    & (\AA) &  (\kms)           \\
\hline
C{\sc iii}   &  1175.71    &   -2.4 $\pm$ 0.8   &    21 $\pm$ 101 \\
H{\sc i}     &  1215.67    &   20.9 $\pm$ 2.9   &   376 $\pm$ 13 \\
Si{\sc ii}   &  1260.42    &   -1.4 $\pm$ 0.3   &  -281 $\pm$ 38 \\
Si{\sc ii}*  &  1264.74    &    0.9 $\pm$ 0.3   &    89 $\pm$ 61 \\
O{\sc i} + Si{\sc ii}  &  1303.27    &   -1.6 $\pm$ 0.3   &  -258 $\pm$ 62 \\
Si{\sc ii}*  &  1309.28    &    0.7 $\pm$ 0.2   &  -118 $\pm$ 80 \\
C{\sc ii}    &  1334.53    &   -1.6 $\pm$ 0.2   &  -136 $\pm$ 39 \\
Si{\sc iv}   &  1393.76    &   -1.1 $\pm$ 0.2   &  -122 $\pm$ 50 \\
Si{\sc iv}   &  1402.77    &   -0.8 $\pm$ 0.2   &  -160 $\pm$ 45 \\
Si{\sc ii}   &  1526.71    &   -1.3 $\pm$ 0.3   &   -99 $\pm$ 54 \\
Si{\sc ii}*  &  1533.43    &    0.8 $\pm$ 0.3   &   114 $\pm$ 64 \\
C{\sc iv}    &  1549.48    &   -2.6 $\pm$ 0.4   &  -374 $\pm$ 57 \\
He{\sc ii}   &  1640.40    &    1.3 $\pm$ 0.7   &   141 $\pm$ 282 \\
\hline
\end{tabular}
\caption{\label{tab:lines} Equivalent width and velocity of absorption and emission lines in the composite spectrum (Figure~\ref{fig:composite}).}
\end{table}

\begin{table}[p]
\begin{tabular}{lccc}
\hline
\hline
Property  &  $z<4.1$  &  $z>4.1$  \\
\hline
$z$               &  $3.76 \pm 0.21$   &  $4.70 \pm 0.58$  \\
$M_{UV}$          &  $-21.0 \pm 0.3$   &  $-21.0 \pm 0.3$  \\
$\log M_*$/\Msun  &  $9.6 \pm 0.6$     &  $9.8 \pm 0.6$    \\
$r_h$ (kpc)       &  $2.42 \pm 0.97$   &  $1.94 \pm 0.74$  \\
$\beta$           &  $-2.02 \pm 0.08$  &  $-2.12 \pm 0.19$ \\
$W_{\Lya}$ (\AA)  &  $15.4 \pm 5.9$    &  $12.2 \pm 3.7$   \\
$W_{1260}$ (\AA)  &  $-1.7 \pm 0.4$    &  $-0.7 \pm 0.4$   \\
$W_{1265}$ (\AA)  &  $0.8 \pm 0.3$     &  $1.0 \pm 1.0$    \\
$W_{1303}$ (\AA)  &  $-2.1 \pm 0.4$    &  $-0.9 \pm 0.4$   \\
$W_{1309}$ (\AA)  &  $1.0 \pm 0.3$     &  $0.9 \pm 0.4$    \\
$W_{1334}$ (\AA)  &  $-1.7 \pm 0.3$    &  $-1.7 \pm 0.9$   \\
$W_{1527}$ (\AA)  &  $-1.2 \pm 0.3$    &  $-0.6 \pm 1.0$   \\
$W_{1533}$ (\AA)  &  $0.7 \pm 0.3$     &  $0.9 \pm 0.5$    \\
\hline
\end{tabular}
\caption{\label{tab:redshift} Mean demographic and spectroscopic properties of LBGs at different redshifts. Error bars correspond to the standard deviation of values for individual galaxies in each sub-sample. $\beta$ and equivalent widths of \Lya and low-ionization metal transitions are measured directly from composite spectra, with error bars determined from a bootstrap method (see text for details).}
\end{table}

\footnotesize


\bibliography{deimos}{}

\begin{thebibliography}{}

\bibitem[Balestra et al.(2010)]{Balestra10} Balestra, I., Mainieri, V., Popesso, P., et al.\ 2010, \aap, 512, A12 

\bibitem[Bouwens et al.(2004)]{Bouwens04} Bouwens, R.~J., Illingworth, G.~D., Blakeslee, J.~P., Broadhurst, T.~J., \& Franx, M.\ 2004, \apjl, 611, L1 

\bibitem[Bouwens et al.(2007)]{Bouwens07} Bouwens, R.~J., Illingworth, G.~D., Franx, M., \& Ford, H.\ 2007, \apj, 670, 928 

\bibitem[Bouwens et al.(2009)]{Bouwens09} Bouwens, R.~J., Illingworth, G.~D., Franx, M., et al.\ 2009, \apj, 705, 936 

\bibitem[Bouwens et al.(2011a)]{Bouwens11a} Bouwens, R.~J., Illingworth, G.~D., Oesch, P.~A., et al.\ 2011, \apj, 737, 90 

\bibitem[Bouwens et al.(2011b)]{Bouwens11b} Bouwens, R.~J., Illingworth, G.~D., Oesch, P.~A., et al.\ 2011, arXiv:1109.0994 

\bibitem[Brammer et al.(2011)]{Brammer11} Brammer, G.~B., Whitaker, K.~E., van Dokkum, P.~G., et al.\ 2011, \apj, 739, 24 

\bibitem[Davis et al.(2003)]{Davis03} Davis, M., Faber, S.~M., Newman, J., et al.\ 2003, \procspie, 4834, 161 

\bibitem[Dekel et al.(2009)]{Dekel09} Dekel, A., Birnboim, Y., Engel, G., et al.\ 2009, \nat, 457, 451 

\bibitem[Eldridge \& Stanway(2009)]{Eldridge09} Eldridge, J.~J., \& Stanway, E.~R.\ 2009, \mnras, 400, 1019 

\bibitem[Eldridge \& Stanway(2011)]{Eldridge11} Eldridge, J.~J., \& Stanway, E.~R.\ 2011, \mnras, 1595 

\bibitem[Ellis(2008)]{Ellis08} Ellis, R.~S.\ 2008, Saas-Fee Advanced Course 36: First Light in the Universe, 259

\bibitem[Erb et al.(2010)]{Erb10} Erb, D.~K., Pettini, M., Shapley, A.~E., et al.\ 2010, \apj, 719, 1168 

\bibitem[Faber et al.(2003)]{Faber03} Faber, S.~M., Phillips, A.~C., Kibrick, R.~I., et al.\ 2003, \procspie, 4841, 1657 

\bibitem[Faucher-Gigu{\`e}re et al.(2011)]{Faucher11} Faucher-Gigu{\`e}re, C.-A., Kere{\v s}, D., \& Ma, C.-P.\ 2011, \mnras, 417, 2982 

\bibitem[Ferguson et al.(2004)]{Ferguson04} Ferguson, H.~C., Dickinson, M., Giavalisco, M., et al.\ 2004, \apjl, 600, L107 

\bibitem[Forero-Romero et al.(2011)]{Forero11} Forero-Romero, J.~E., Yepes, G., Gottloeber, S., \& Prada, F.\ 2011, arXiv:1109.0228 

\bibitem[Giavalisco et al.(2004)]{Giavalisco04} Giavalisco, M., Ferguson, H.~C., Koekemoer, A.~M., et al.\ 2004, \apjl, 600, L93 

\bibitem[Gonz{\'a}lez et al.(2011)]{Gonzalez11} Gonz{\'a}lez, V., Labb{\'e}, I., Bouwens, R.~J., et al.\ 2011, \apjl, 735, L34 

\bibitem[Grogin et al.(2011)]{Grogin11} Grogin, N.~A., Kocevski, D.~D., Faber, S.~M., et al.\ 2011, arXiv:1105.3753 

\bibitem[Haiman \& Spaans(1999)]{Haiman99} Haiman, Z., \& Spaans, M.\ 1999, \apj, 518, 138 

\bibitem[Heckman(2002)]{Heckman02} Heckman, T.~M.\ 2002, Extragalactic Gas at Low Redshift, 254, 292 

\bibitem[Hopkins \& Beacom(2006)]{Hopkins06} Hopkins, A.~M., \& Beacom, J.~F.\ 2006, \apj, 651, 142 

\bibitem[Koekemoer et al.(2011)]{Koekemoer11} Koekemoer, A.~M., Faber, S.~M., Ferguson, H.~C., et al.\ 2011, arXiv:1105.3754 

\bibitem[Kobulnicky \& Skillman(1998)]{Kobulnicky98} Kobulnicky, H.~A., \& Skillman, E.~D.\ 1998, \apj, 497, 601 

\bibitem[Kroupa(2002)]{Kroupa02} Kroupa, P.\ 2002, Science, 295, 82 

\bibitem[Maiolino et al.(2008)]{Maiolino08} Maiolino, R., Nagao, T., Grazian, A., et al.\ 2008, \aap, 488, 463 

\bibitem[Masters \& Capak(2011)]{Masters11} Masters, D., \& Capak, P.\ 2011, \pasp, 123, 638 

\bibitem[Murray et al.(2010)]{Murray10} Murray, N., Quataert, E., \& Thompson, T.~A.\ 2010, \apj, 709, 191 

\bibitem[Oke(1974)]{Oke74} Oke, J.~B.\ 1974, \apjs, 27, 21 

\bibitem[Ono et al.(2011)]{Ono11} Ono, Y., Ouchi, M., Mobasher, B., et al.\ 2011, arXiv:1107.3159 

\bibitem[Ouchi et al.(2008)]{Ouchi08} Ouchi, M., Shimasaku, K., Akiyama, M., et al.\ 2008, \apjs, 176, 301 

\bibitem[Pentericci et al.(2011)]{Pentericci11} Pentericci, L., Fontana, A., Vanzella, E., et al.\ 2011, arXiv:1107.1376 

\bibitem[Pettini et al.(2000)]{Pettini00} Pettini, M., Steidel, C.~C., Adelberger, K.~L., Dickinson, M., \& Giavalisco, M.\ 2000, \apj, 528, 96 

\bibitem[Pettini et al.(2002)]{Pettini02} Pettini, M., Rix, S.~A., Steidel, C.~C., et al.\ 2002, \apj, 569, 742 

\bibitem[Quider et al.(2009)]{Quider09} Quider, A.~M., Pettini, M., Shapley, A.~E., \& Steidel, C.~C.\ 2009, \mnras, 398, 1263 

\bibitem[Quider et al.(2010)]{Quider10} Quider, A.~M., Shapley, A.~E., Pettini, M., Steidel, C.~C., \& Stark, D.~P.\ 2010, \mnras, 402, 1467 

\bibitem[Reddy \& Steidel(2009)]{Reddy09} Reddy, N.~A., \& Steidel, C.~C.\ 2009, \apj, 692, 778 

\bibitem[Robertson et al.(2010)]{Robertson10} Robertson, B.~E., Ellis, R.~S., Dunlop, J.~S., McLure, R.~J., \& Stark, D.~P.\ 2010, \nat, 468, 49 

\bibitem[Schenker et al.(2011)]{Schenker11} Schenker, M.~A., Stark, D.~P, Ellis, R.~S., et al.\ 2011, arXiv:1107.1261 

\bibitem[Shapley et al.(2001)]{Shapley01} Shapley, A.~E., Steidel, C.~C., Adelberger, K.~L., et al.\ 2001, \apj, 562, 95 

\bibitem[Shapley et al.(2003)]{Shapley03} Shapley, A.~E., Steidel, C.~C., Pettini, M., \& Adelberger, K.~L.\ 2003, \apj, 588, 65 

\bibitem[Shapley(2011)]{Shapley11} Shapley, A.~E.\ 2011, \araa, 49, 525 

\bibitem[Stark et al.(2009)]{Stark09} Stark, D.~P., Ellis, R.~S., Bunker, A., et al.\ 2009, \apj, 697, 1493 

\bibitem[Stark et al.(2010)]{Stark10} Stark, D.~P., Ellis, R.~S., Chiu, K., Ouchi, M., \& Bunker, A.\ 2010, \mnras, 408, 1628 

\bibitem[Stark et al.(2011)]{Stark11} Stark, D.~P., Ellis, R.~S., \& Ouchi, M.\ 2011, \apjl, 728, L2 

\bibitem[Steidel et al.(2003)]{Steidel03} Steidel, C.~C., Adelberger, K.~L., Shapley, A.~E., et al.\ 2003, \apj, 592, 728 

\bibitem[Steidel et al.(2010)]{Steidel10} Steidel, C.~C., Erb, D.~K., Shapley, A.~E., et al.\ 2010, \apj, 717, 289 

\bibitem[Steidel et al.(2011)]{Steidel11} Steidel, C.~C., Bogosavljevi{\'c}, M., Shapley, A.~E., et al.\ 2011, \apj, 736, 160 

\bibitem[Vanzella et al.(2005)]{Vanzella05} Vanzella, E., Cristiani, S., Dickinson, M., et al.\ 2005, \aap, 434, 53 

\bibitem[Vanzella et al.(2006)]{Vanzella06} Vanzella, E., Cristiani, S., Dickinson, M., et al.\ 2006, \aap, 454, 423 

\bibitem[Vanzella et al.(2008)]{Vanzella08} Vanzella, E., Cristiani, S., Dickinson, M., et al.\ 2008, \aap, 478, 83 

\bibitem[Vanzella et al.(2009)]{Vanzella09} Vanzella, E., Giavalisco, M., Dickinson, M., et al.\ 2009, \apj, 695, 1163 

\bibitem[Wang et al.(2010)]{Wang10} Wang, W.-H., Cowie, L.~L., Barger, A.~J., Keenan, R.~C., \& Ting, H.-C.\ 2010, \apjs, 187, 251 

\bibitem[Wardlow et al.(2011)]{Wardlow11} Wardlow, J.~L., Smail, I., Coppin, K.~E.~K., et al.\ 2011, \mnras, 415, 1479 

\bibitem[Weiner et al.(2009)]{Weiner09} Weiner, B.~J., Coil, A.~L., Prochaska, J.~X., et al.\ 2009, \apj, 692, 187 

\end{thebibliography}
\bibliographystyle{apj}

\end{document}